\documentclass[%
reprint,
superscriptaddress,
 amsmath,amssymb,
 aps,
]{revtex4-2}

\usepackage{graphicx}
\usepackage{dcolumn}
\usepackage{bm}

\usepackage[caption=false]{subfig}
\usepackage{siunitx, soul, xcolor}
\usepackage{adjustbox}
\usepackage{subfig}
\usepackage{graphicx}
\usepackage[normalem]{ulem}

\begin{document}
\preprint{APS/123-QED}

\title{Room-Temperature Magnetic Skyrmions in Pt/Co/Cu Multilayers}

\author{Shuyu Cheng}
\affiliation{Department of Physics, The Ohio State University, Columbus, Ohio 43210, United States}
\author{N\'uria Bagu\'es}
\affiliation{Department of Materials Science and Engineering, The Ohio State University, Columbus, Ohio 43210, United States}
\author{Camelia M. Selcu}
\email{selcu.1@osu.edu}
\affiliation{Department of Physics, The Ohio State University, Columbus, Ohio 43210, United States}
\author{Jacob B. Freyermuth}
\affiliation{Department of Physics, The Ohio State University, Columbus, Ohio 43210, United States}
\author{Ziling Li}
\affiliation{Department of Physics, The Ohio State University, Columbus, Ohio 43210, United States}
\author{Binbin Wang}
\affiliation{Department of Materials Science and Engineering, The Ohio State University, Columbus, Ohio 43210, United States}
\author{Shekhar Das}
\affiliation{Department of Physics, The Ohio State University, Columbus, Ohio 43210, United States}
\author{P. Chris Hammel}
\affiliation{Department of Physics, The Ohio State University, Columbus, Ohio 43210, United States}
\author{Mohit Randeria}
\affiliation{Department of Physics, The Ohio State University, Columbus, Ohio 43210, United States}
\author{David W. McComb}
\email{mccomb.29@osu.edu}
\affiliation{Department of Materials Science and Engineering, The Ohio State University, Columbus, Ohio 43210, United States}
\author{Roland K. Kawakami}
\email{kawakami.15@osu.edu}
\affiliation{Department of Physics, The Ohio State University, Columbus, Ohio 43210, United States}

\begin{abstract}

Magnetic skyrmions are promising for next-generation information storage and processing owing to their potential advantages in data storage density, robustness, and energy efficiency.
The magnetic multilayers consisting of Pt, Co, and a third metal element $X$ provide an ideal platform to study the skyrmions due to their highly tunable magnetic properties.
Here, we report the observation of room-temperature bubble-like N\'eel skyrmions in epitaxial Pt/Co/Cu multilayers in samples with multidomain states in zero field.
The magneto-optic Kerr effect (MOKE) and superconducting quantum interference device (SQUID) magnetometry are applied to investigate the shapes of the hysteresis loops, the magnetic anisotropy, and the saturation magnetization.
By tuning the Co thickness and the number of periods, we achieve perpendicular and in-plane magnetized states and multidomain states that are identified by a wasp-waisted hysteresis loop.
Skyrmions are directly imaged by magnetic force microscopy (MFM) and Lorentz transmission electron microscopy (LTEM).
The development of room-temperature skyrmions in Pt/Co/Cu multilayers may lead to advances in skyrmion-related research and applications.

\end{abstract}

\flushbottom
\maketitle
\thispagestyle{empty}

\begin{figure*}[ht]
  
    \subfloat
    [\label{fig:structure}]{\includegraphics[width=0.38\textwidth]{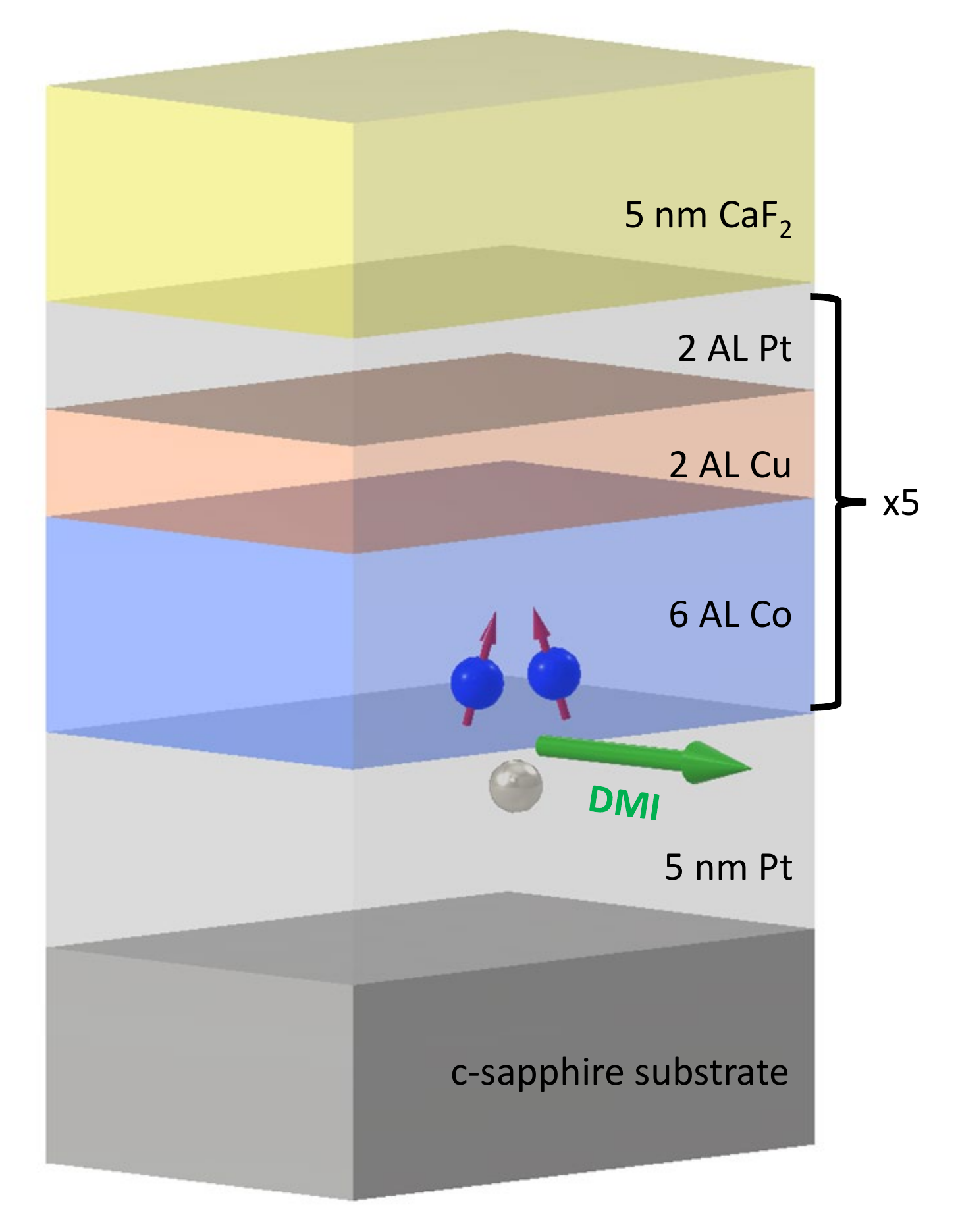}
    }
    \subfloat
    [\label{fig:RHEED}]{\includegraphics[width=0.26\textwidth]{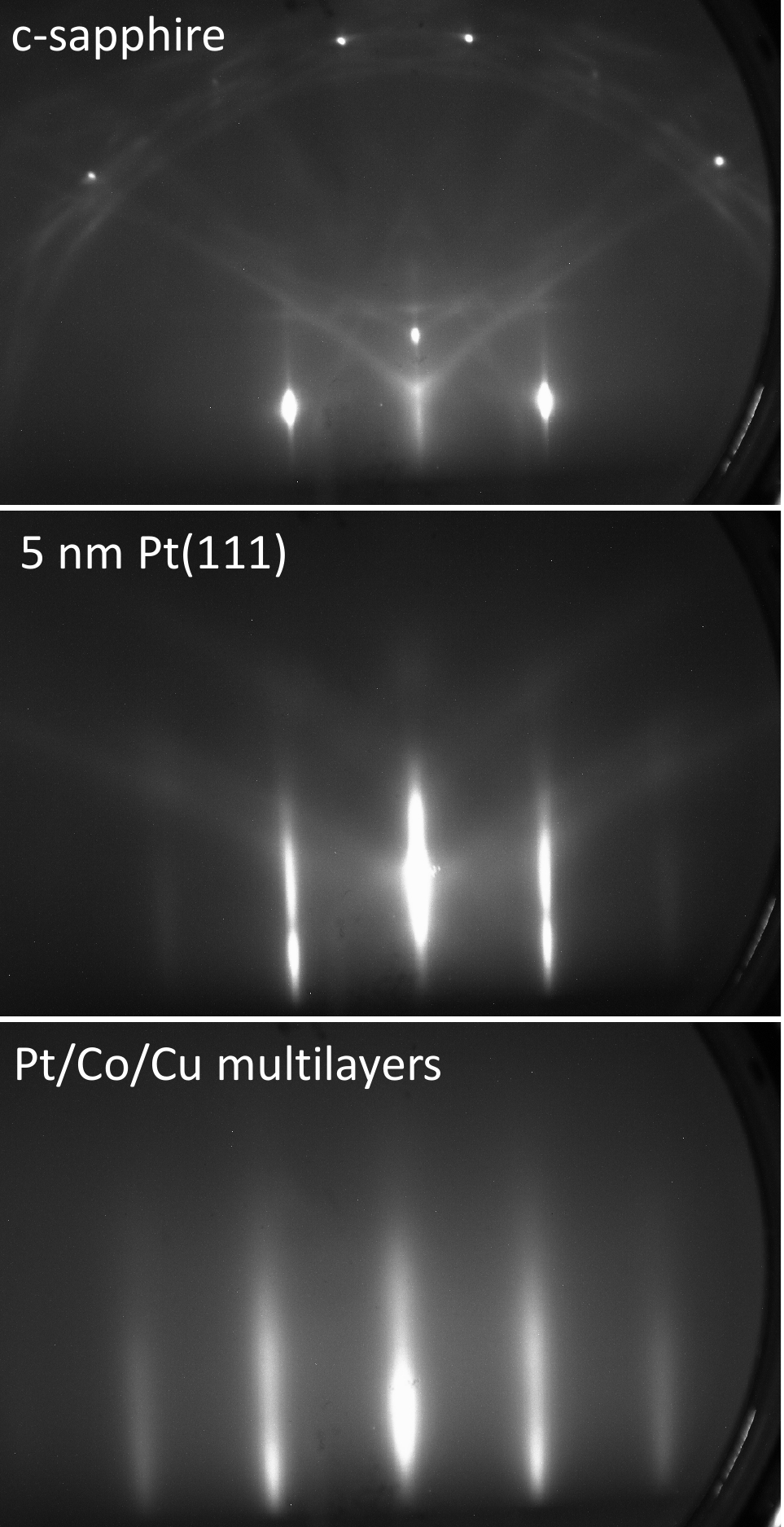}
   }
  \begin{minipage}[b]{.32\textwidth}
   \subfloat[\label{fig:RHEED_Osc}]{
       \includegraphics[width=\textwidth]{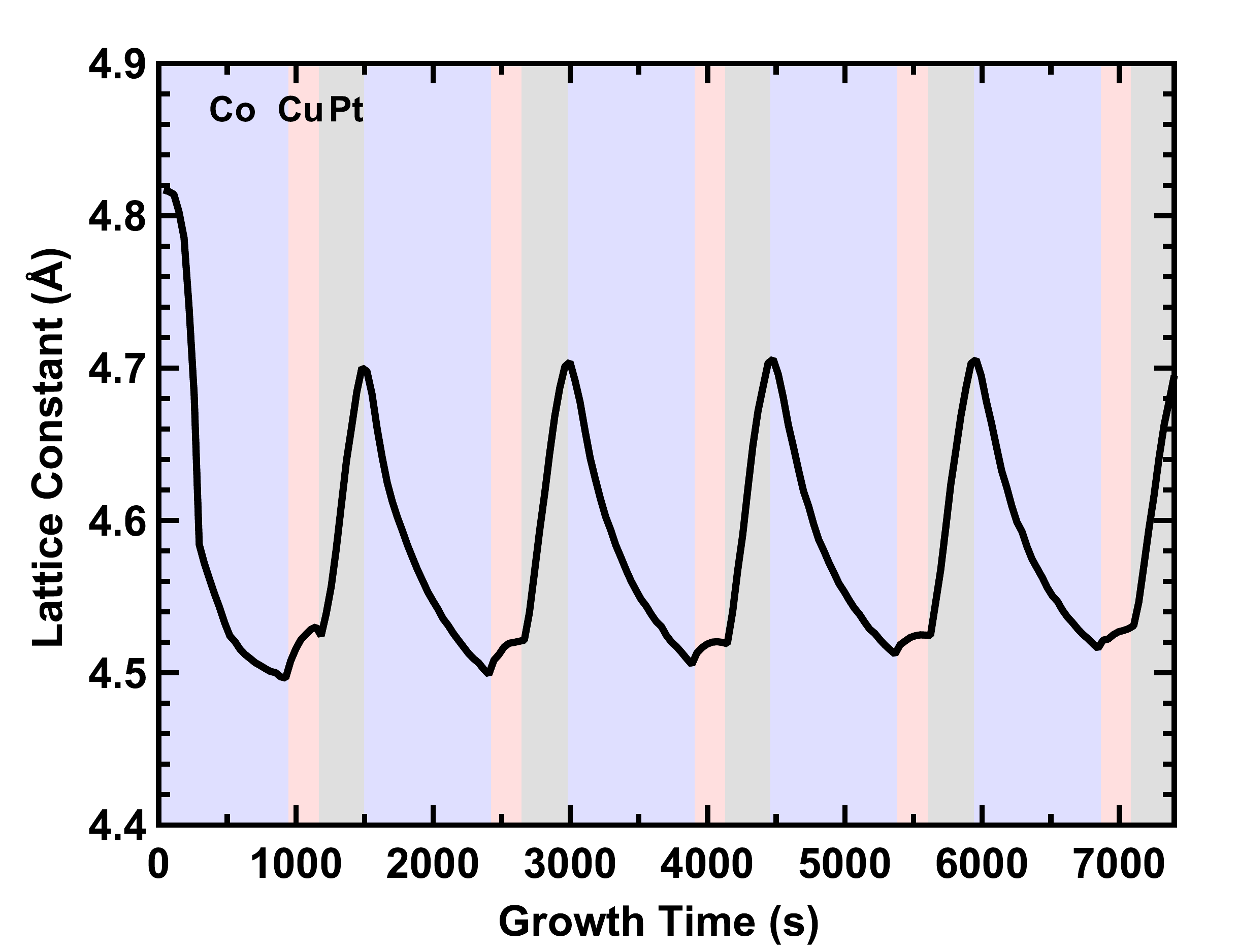}
  }\vfill
  \subfloat[\label{fig:TEM}]{
       \includegraphics[width=0.9\textwidth]{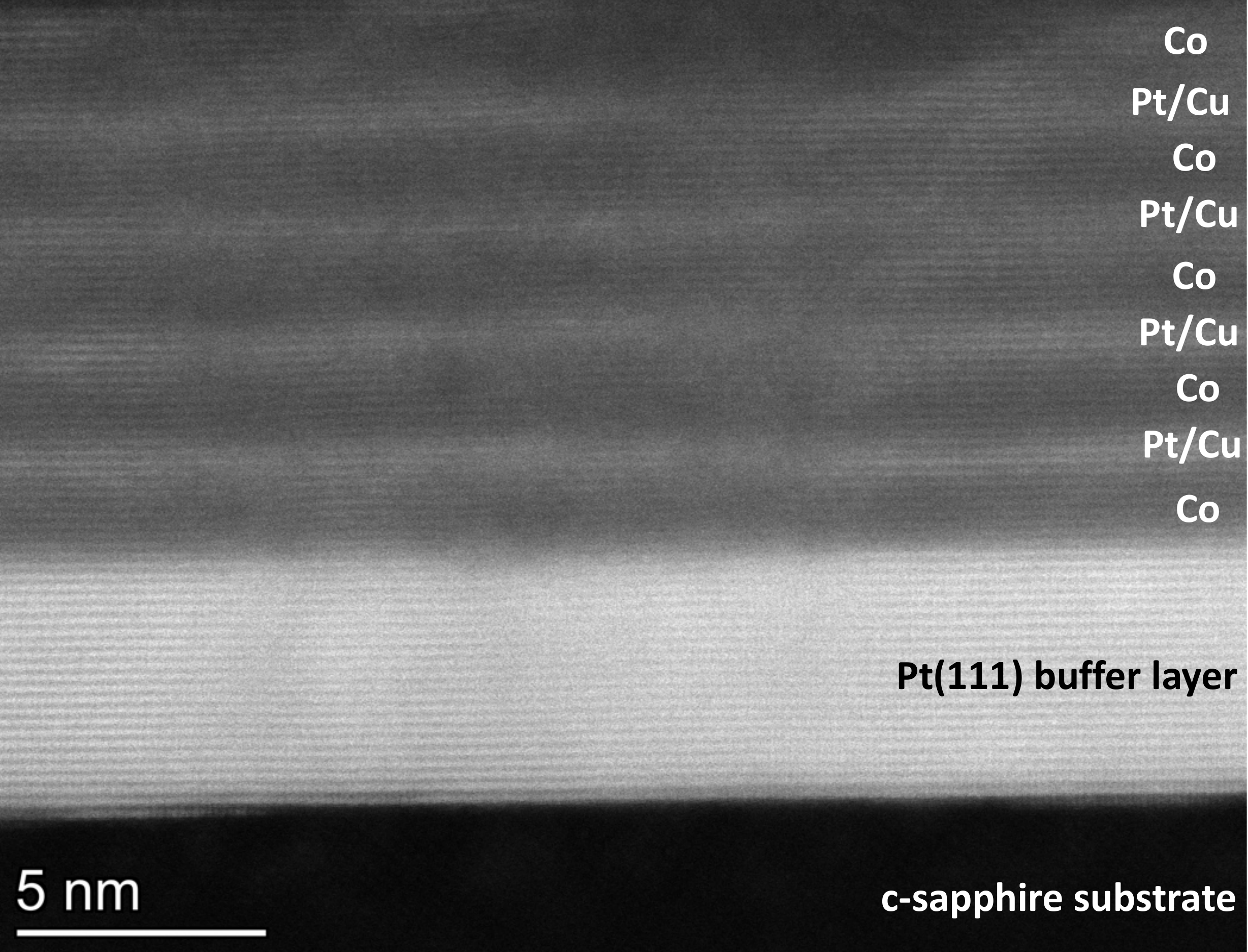}
    } 
  \end{minipage}
  \caption{Material growth and structural characterization. 
    (a) Schematic drawing of the sample structure. 
    (b) \textit{In situ} RHEED patterns during growth. Top: sapphire(0001) with the beam along the [11$\bar{2}$0] in-plane direction. Middle: Pt(111). Bottom: [Pt(2)/Co(6)/Cu(2)]$_5$ multilayers.
    (c) In-plane lattice constant extracted from the RHEED streak spacing during the growth. The blue, red, and gray regions correspond to the deposition of Co, Cu, and Pt, respectively. 
    (d) STEM-HAADF image of [Pt(2)/Co(6)/Cu(2)]$_5$ multilayers.
    }
\end{figure*}

Magnetic skyrmions are topologically protected spin textures that stand out as one of the strongest candidates for next-generation information storage due to their small sizes, thermal stability, and high energy efficiency
~\cite{fert2013skyrmions, koshibae2015memory, finocchio2016magnetic, back20202020}.
These special spin textures originate from the complex interplay between exchange stiffness, magnetic anisotropy, dipolar interactions, applied field, and the Dzyaloshinskii-Moriya interaction (DMI).
The DMI plays an important role in the stabilization of skyrmions, as it favors perpendicular alignment between neighboring spins \cite{dzyaloshinsky1958thermodynamic, moriya1960anisotropic}. 
From a material point of view, the DMI is allowed by asymmetric crystal structures, which happens in bulk crystals that are non-centrosymmetric and at interfaces due to the broken mirror plane symmetry~\cite{wiesendanger2016nanoscale}.
While the former gives rise to Bloch skyrmions in which the spins twist in the tangential direction~\cite{Muhlbauer2009skyrmion, yu2010real}, the latter gives rise to N\'eel skyrmions in which the spins tumble in the radial direction~\cite{moreau2016additive, woo2016observation}. 
Though less common, N\'eel skyrmions could also form in bulk crystals when a mirror plane symmetry is broken~\cite{kezsmarki2015neel}.

Since 2009, magnetic skyrmions have been discovered in a variety of materials, including B20-phase materials~\cite{Muhlbauer2009skyrmion, yu2010real, yu2011near}, two-dimensional materials~\cite{ding2019observation, wu2020neel}, and magnetic bilayers and multilayers~\cite{romming2015field, moreau2016additive, woo2016observation, jiang2017skyrmions}.
Among these materials, the Pt/Co/$X$ ($X$ = metallic material) magnetic multilayers have drawn much attention because the insertion of the $X$ layers into Pt/Co superlattices generates non-canceling interfacial DMI by breaking the inversion symmetry~\cite{moreau2016additive, mcvitie2018transmission, schlotter2018temperature}.
Furthermore, the magnetic properties of Pt/Co/$X$ multilayers can be vastly tuned through varying the thickness of each layer~\cite{sun2016magnetic, bandiera2013effect} and the number of repetitions of Pt/Co/$X$~\cite{benguettat2019interfacial, jena2021interfacial}, or simply changing the element $X$~\cite{belmeguenai2019influence, ajejas2021element}.
So far, the magnetic properties of Pt/Co/$X$ multilayer systems with various metallic materials $X$ has been reported including $X$ = Mn~\cite{lonsky2022structural}, Ni~\cite{rojas2016perpendicular}, Cu~\cite{sun2016magnetic}, Ru~\cite{karayev2019interlayer}, Ho~\cite{liu2020influence}, Ta~\cite{woo2016observation}, W~\cite{benguettat2019interfacial}, Ir~\cite{moreau2016additive}, etc.
Within a number of options for transition metal $X$, Cu is of particular interest for a number of reasons.
Since the lattice constant of Cu is close to that of Co and Pt, it is possible to grow Pt/Co/Cu multilayers epitaxially along the Pt(111) direction~\cite{sun2016magnetic}.
This enables the layer-by-layer growth of high-quality crystalline Pt/Co/Cu multilayers using molecular beam epitaxy (MBE).
Previous studies also reported that the Pt/Co/Cu multilayers have no magnetic dead layer as compared to other materials as $X$ layer~\cite{belmeguenai2019influence}. 
For these reasons, Pt/Co/Cu could be a model system to investigate skyrmion properties.

In this paper, we report the observation of room-temperature bubble-like N\'eel skyrmions in epitaxial Pt/Co/Cu multilayers, which was achieved by applying an out-of-plane (OOP) magnetic field to samples with a zero-field multidomain state.
We used magneto-optic Kerr effect (MOKE) and magnetic force microscopy (MFM) to investigate how the zero-field magnetic states depend on the Co thickness ($t_{Co}$) and number of periods $N$.
We observed a spin-reorientation transition from OOP-magnetized to multidomain to in-plane (IP) magnetized with increasing $t_{Co}$, and a transition from OOP-magnetized to multidomain with increasing $N$. 
The multidomain states of interest could be recognized by the wasp-waisted shapes their OOP MOKE loops, and applying an OOP field generated skyrmions as observed by Lorentz transmission electron microscopy (LTEM) and MFM. 
In addition, since we developed the growth of epitaxial [Pt/Co/Cu]$_N$ multilayers on an insulating substrate, we were able to apply current pulses which assisted in the nucleation of skyrmions. To better understand the properties of the skyrmions and the multidomain states, we employed micromagnetic simulations which were able to reproduce the skyrmion size, the domain size in the multidomain state, and the transition from OOP-magnetized to multidomain with increasing $N$.

\begin{figure*}[ht]

    \subfloat[\label{fig:MOKE}]{
    \includegraphics[width=0.48\textwidth]{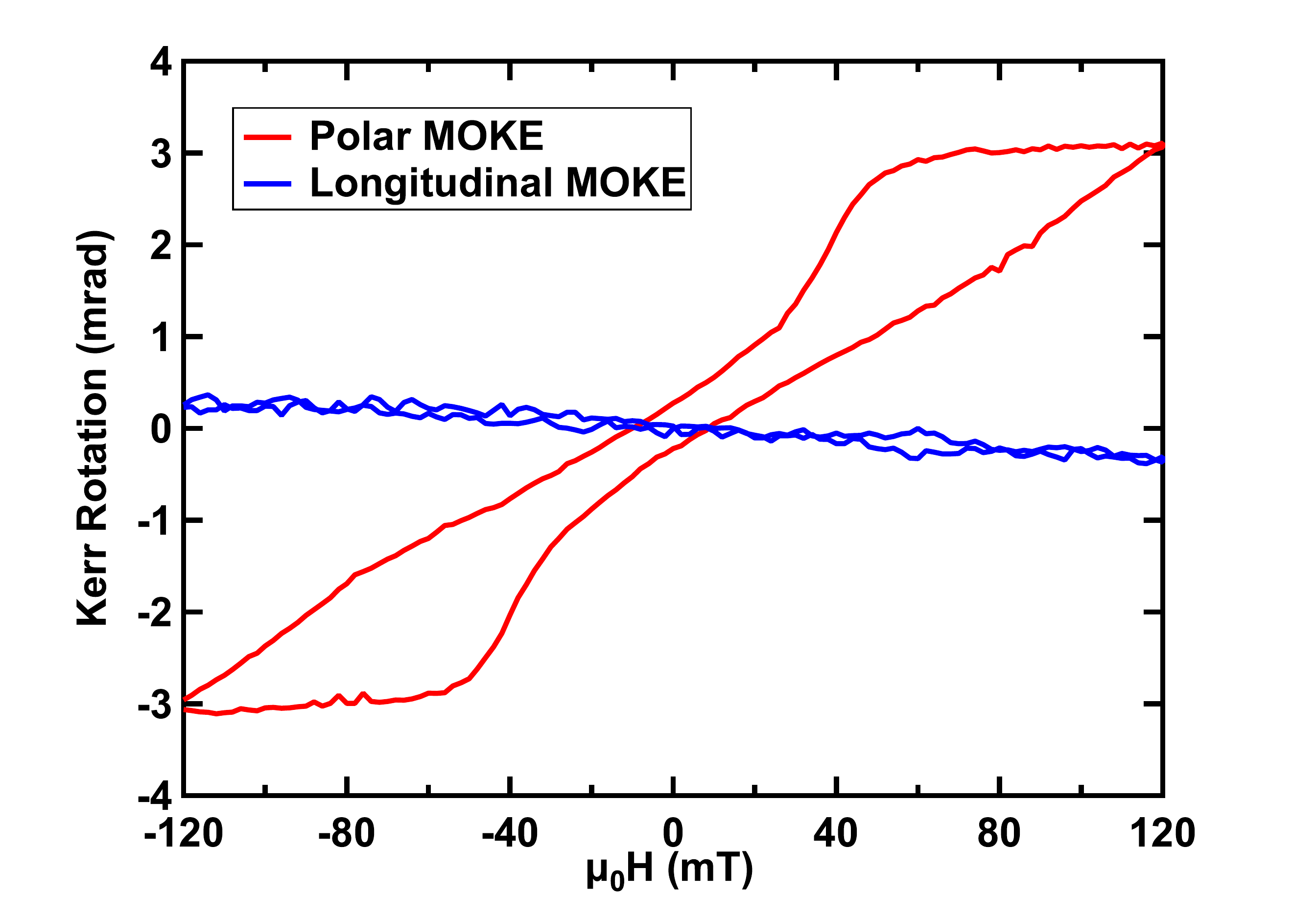}
    }\hfill
    \subfloat[\label{fig:SQUID}]{
       \includegraphics[width=0.48\textwidth]{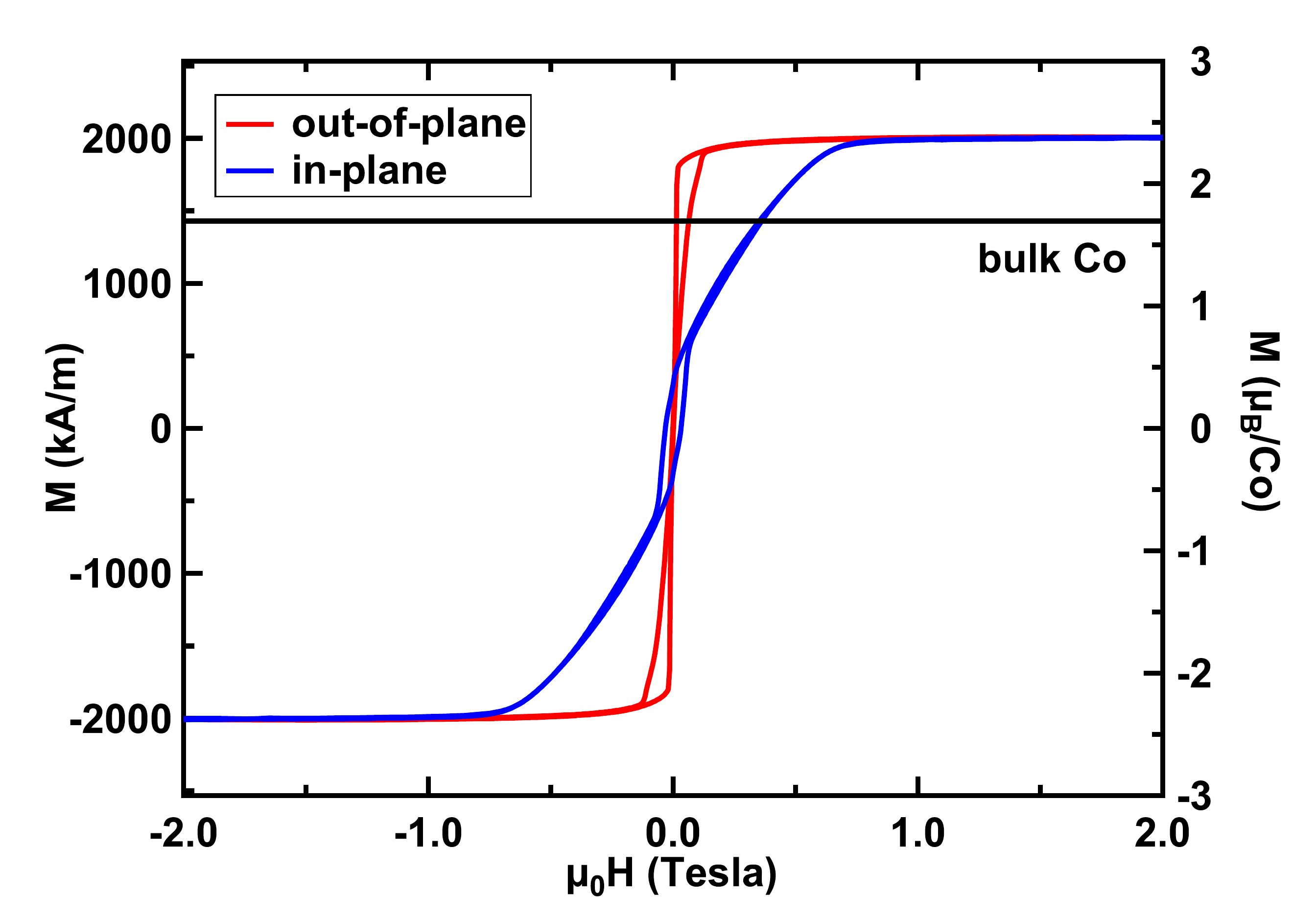}
    }
    \caption{\label{fig:Magnetic} Magnetic characterizations of [Pt(2)/Co(6)/Cu(2)]$_5$ multilayers.
    (a) MOKE hysteresis loops of [Pt(2)/Co(6)/Cu(2)]$_5$ multilayers.
    (b) SQUID hysteresis loops of [Pt(2)/Co(6)/Cu(2)]$_5$ multilayers. The black solid line represents 1430\,kA/m, which is the saturation magnetization of bulk Co. 
    } 
\end{figure*}

The [Pt/Co/Cu]$_N$ multilayers were grown on epitaxial Pt(111) buffer layers on Al$_2$O$_3$(0001) substrates (MTI Corporation) using MBE. 
Unless otherwise noted, the structure of the multilayer samples is (bottom to top): 5\,nm Pt/[6 atomic layers (6\,AL) Co/2\,AL Cu/2\,AL Pt]$_5$/5\,nm CaF$_2$ (hereafter [Pt(2)/Co(6)/Cu(2)]$_5$, 
where the sequence of the layers is from the bottom to the top, and the numbers in the parentheses represent the thickness of each layer in units of atomic layers), as shown in Figure~\ref{fig:structure}.
Prior to the growth, the Al$_2$O$_3$(0001) substrates were annealed in air at 1000\,$^\circ$C for 180 minutes and then degassed in the growth chamber at 500\,$^\circ$C for 30 minutes.
A 5\,nm Pt(111) buffer layer was epitaxially grown on the Al$_2$O$_3$(0001) substrate following the recipe described in \cite{cheng2022atomic}.
[Pt/Co/Cu]$_N$ multilayers were deposited on top of the Pt(111) buffer layer at room temperature (RT) by opening and closing the shutters sequentially.
The growth time for each layer was determined by the growth rate which was calibrated by a quartz crystal deposition monitor.
Pt was deposited from an electron-beam evaporator, while Co and Cu were deposited from Knudsen cells.
The typical growth rates for Pt, Co, and Cu are 0.9\,\AA/min, 0.8\,\AA/min, and 1.0\,\AA/min, respectively.
After growth, 5\,nm CaF$_2$ was deposited on the sample to protect the sample from oxidation.
The \textit{in situ} reflection high-energy electron diffraction (RHEED) pattern was monitored during the growth, as shown in Figure~\ref{fig:RHEED}.
Streaky RHEED patterns indicate that the [Pt(2)/Co(6)/Co(2)]$_5$ multilayers grow epitaxially. 
Furthermore, the in-plane lattice constant extracted from the RHEED pattern during growth shows oscillatory behavior, with decreasing lattice constant during Co layer growth, and increasing lattice constant during Cu and Pt layer growth, as shown in Figure~\ref{fig:RHEED_Osc}.
This oscillation of in-plane lattice constant does not decay during the growth.

The structure of the sample was confirmed by scanning transmission electron microscopy (STEM) imaging of a cross-section sample using a ThermoFisher probe corrected Themis-Z at 300\,kV. 
The cross-section sample was prepared by Ga ion milling at 30\,kV and 5\,kV using a focused ion beam (FIB). 
Figure~\ref{fig:TEM} shows a STEM high-angle annular dark field (HAADF) image of [Pt(2)/Co(6)/Co(2)]$_5$ multilayer on top of Pt buffer layer on Al$_2$O$_3$ viewed along the Al$_2$O$_3$-[1$\bar{1}$00] axis.
In Figure~\ref{fig:TEM}, the [Pt(2)/Co(6)/Co(2)]$_5$ sample exhibits have well-defined layered structures. 
Due to atomic number (Z-) contrast in HAADF images the Co layers appear as dark layers in the STEM image, while the Pt and Cu layers appear as bright layers.

We first discuss MOKE and SQUID measurements of [Pt(2)/Co(6)/Cu(2)]$_5$ multilayers. 
The MOKE measurement utilized a 632.8 nm laser and employed normal incidence for polar loops and a 45$^\circ$ angle of incidence for longitudinal loops.
Figure~\ref{fig:MOKE} shows the polar (red curve) and longitudinal (blue curve) MOKE hysteresis loops of a [Pt(2)/Co(6)/Cu(2)]$_5$ multilayer.
The polar MOKE hysteresis loop shows a ``wasp-waisted'' shape with small remanence.
Although the applied magnetic field was limited to 120\,mT which is not sufficient to fully saturate the sample, the hysteresis loop nevertheless captures the main magnetic characteristics of the sample.
Fully saturated loops were measured by SQUID (Figure 2b) and are discussed below.
The wasp-waisted shape of the hysteresis loop is similar to that of Pt/Co/Cu multilayers near the spin-reorientation transition (SRT), where magnetic stripe domains with OOP easy axis were observed~\cite{sun2016magnetic}. 
Meanwhile, the longitudinal MOKE hysteresis loop shows almost linear behavior in 120\,mT range, with a much smaller magnitude compared to polar MOKE.
The response to external magnetic fields where the magnetization is easier to polarize out-of-plane compared to in-plane is due to the presence of perpendicular surface magnetic anisotropy from the Pt/Co interfaces.

The results from SQUID measurements of [Pt(2)/Co(6)/Cu(2)]$_5$ are shown in Figure~\ref{fig:SQUID}.
The out-of-plane hysteresis loop (red curve) shows a similar shape to the polar MOKE loop, with low remanence and a saturation field of 192\,mT, while the in-plane hysteresis loop (blue curve) saturates at $\sim$0.7\,Tesla.
We note that the saturation magnetization of [Pt(2)/Co(6)/Cu(2)]$_5$ multilayers is measured to be $2010\pm200$\,kA/m, where the uncertainty is mainly from the uncertainty in film thickness ($\sim$10\%). 
This is larger than the bulk Co value (1430\,kA/m)~\cite{billas1994magnetism} by the amount of $\Delta M_{Co}$ = $580\pm200$\,kA/m.
The additional magnetic moment comes from Pt and is induced by the magnetic proximity effect, which has been reported in Pt/Co multilayers~\cite{bersweiler2016impact}.
To quantify this effect, we calculate the magnetic moment of Pt from the following formula~\cite{bersweiler2016impact}:
\begin{equation}
\Delta M_{Co}\,t_{Co}^{total} = M_{Pt}\,t_{Pt}^{total} \label{eq1}
\end{equation}
Since most of the Pt moment concentrate in the first two atomic layers of Pt adjacent to the Co layers~\cite{mukhopadhyay2020asymmetric}, here we take the thickness of 10 atomic layers of Pt, i.e. 2.3\,nm, as $t_{Pt}^{total}$.
By substituting $t_{Co}^{total}$ = $6.1\pm0.6$\,nm and $t_{Pt}^{total}$ = $2.3\pm0.2$\,nm into Eq.~\ref{eq1}, we get $M_{Pt}$ = $2.6\pm0.7\,\mu_{B}/$Pt. 

\begin{table*}
\begin{tabular*}{\textwidth}{c @{\extracolsep{\fill}} cccc}
 Sample ID & Co thickness t$_{Co}$ & Number of periods N & Zero-field magnetic state \\\hline
 I & 4 & 5 & OOP \\
 II & 5 & 5 & OOP \\
 III & 6 & 5 & Multidomain \\
 IV & 7 & 3 & OOP \\
 V & 7 & 4 & OOP \\
 VI & 7 & 5 & Multidomain \\
 VII & 7 & 7 & Multidomain \\
 VIII & 8 & 4 & Multidomain \\
 IX & 8 & 5 & IP \\
 X & 9 & 5 & IP
\end{tabular*}
\caption{\label{tab:table1} Summary of the structures and zero-field magnetic states of the samples.}
\end{table*}

\begin{figure*}[ht]
   \begin{minipage}[b]{.36\textwidth}
    \subfloat[\label{fig:MOKE_Structural_N5}]{
       \includegraphics[width=\textwidth]{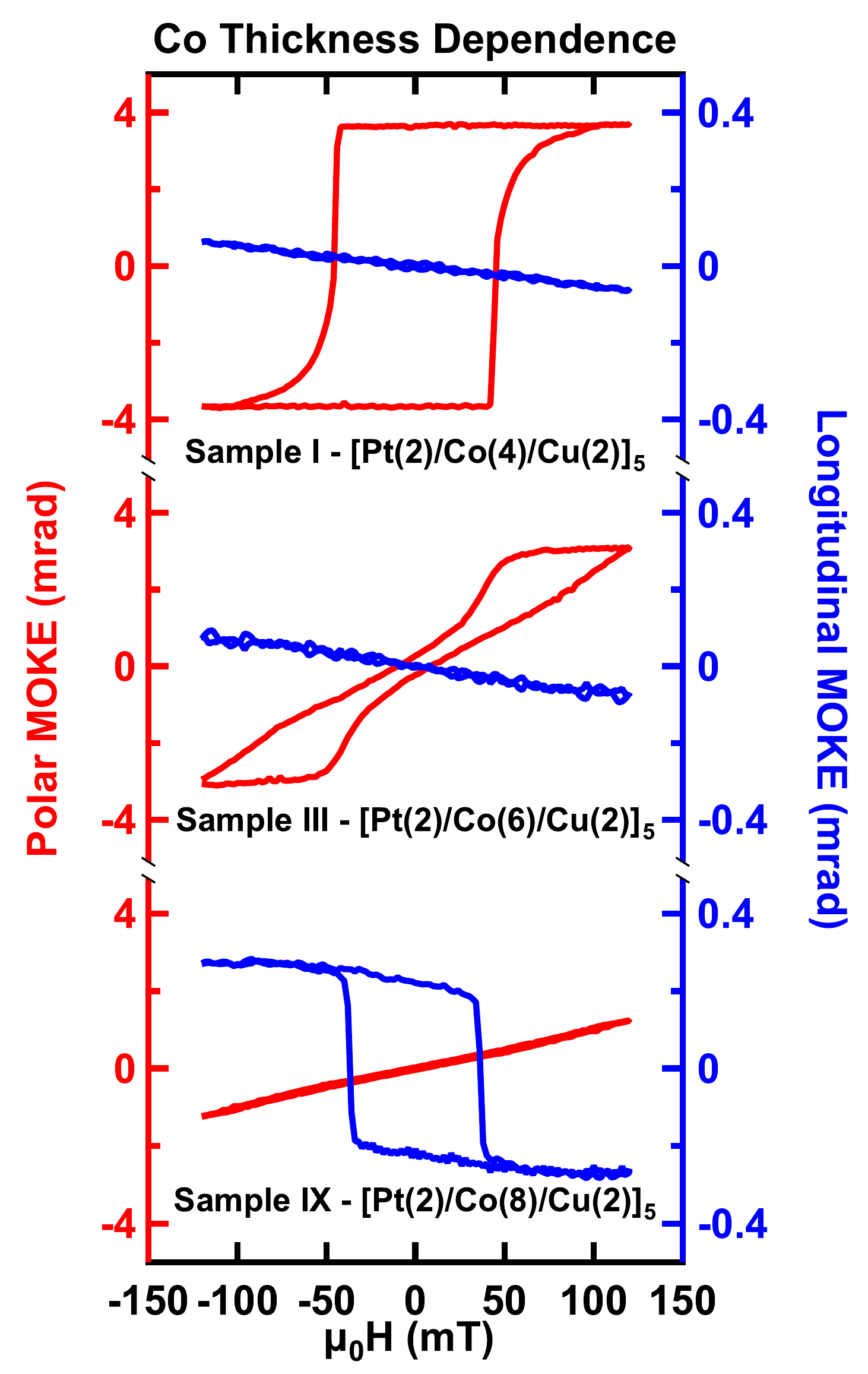}
    }
    \end{minipage}
    \begin{minipage}[b]{0.3\textwidth}
    \subfloat[\label{fig:MOKE_Structural_7ALCo}]{
       \includegraphics[width=\textwidth]{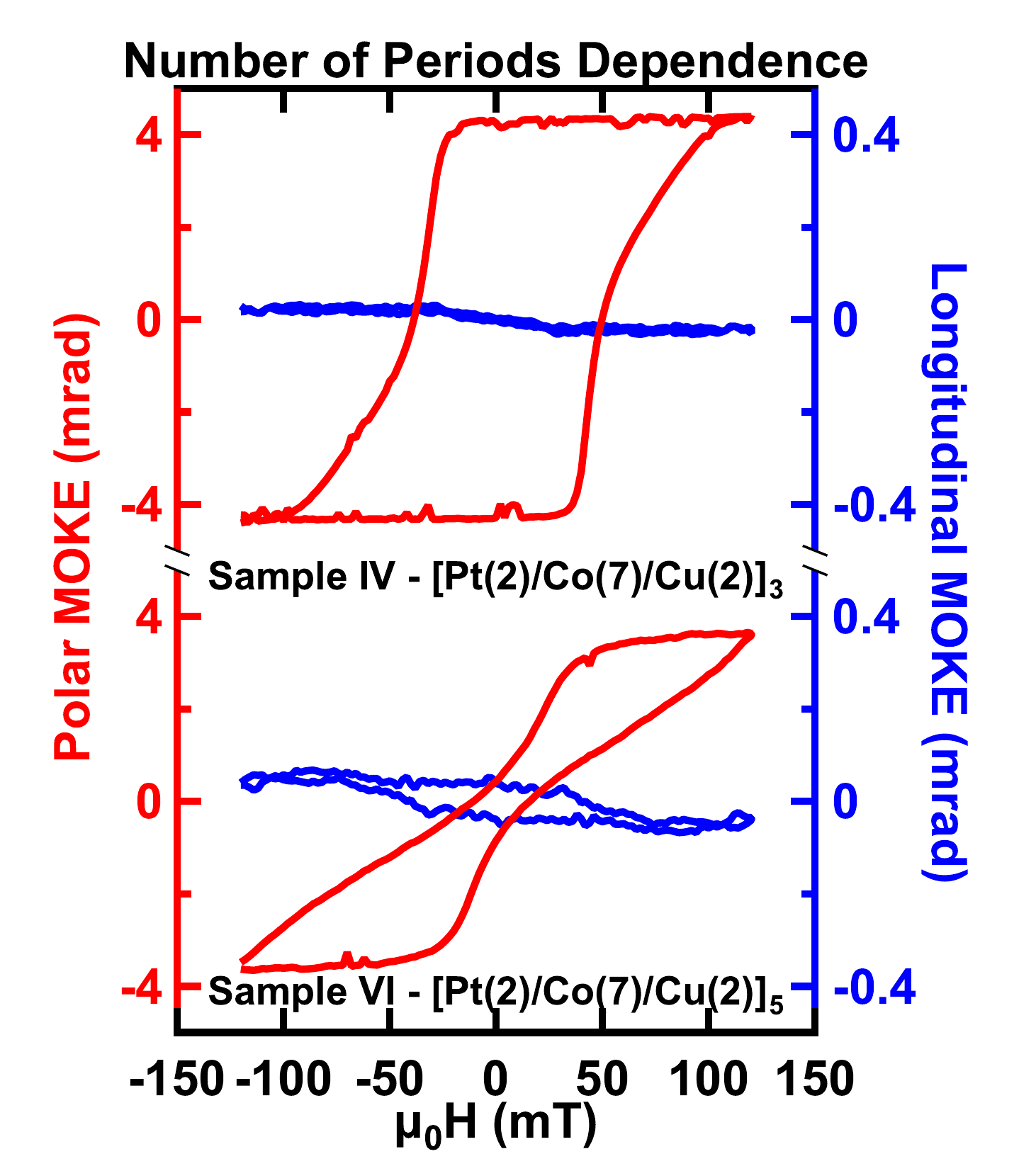}
    }\hfill
    \subfloat[\label{fig:CD/722_7}]{
       \includegraphics[width=\textwidth]{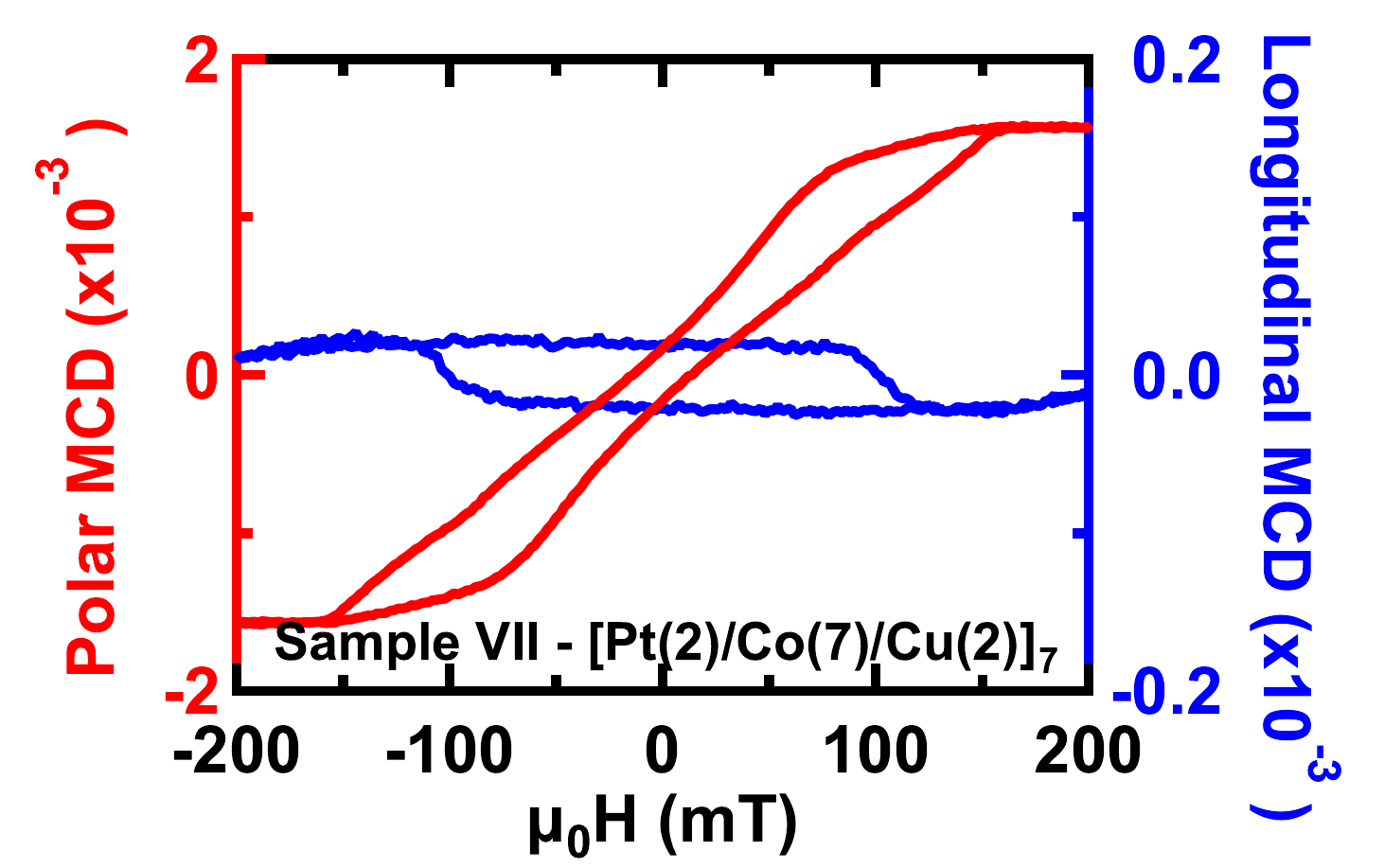}
    }
    \end{minipage}
    \begin{minipage}[b]{.3\textwidth}
    \subfloat[\label{fig:MOKE_20220516}]{
       \includegraphics[width=\textwidth]{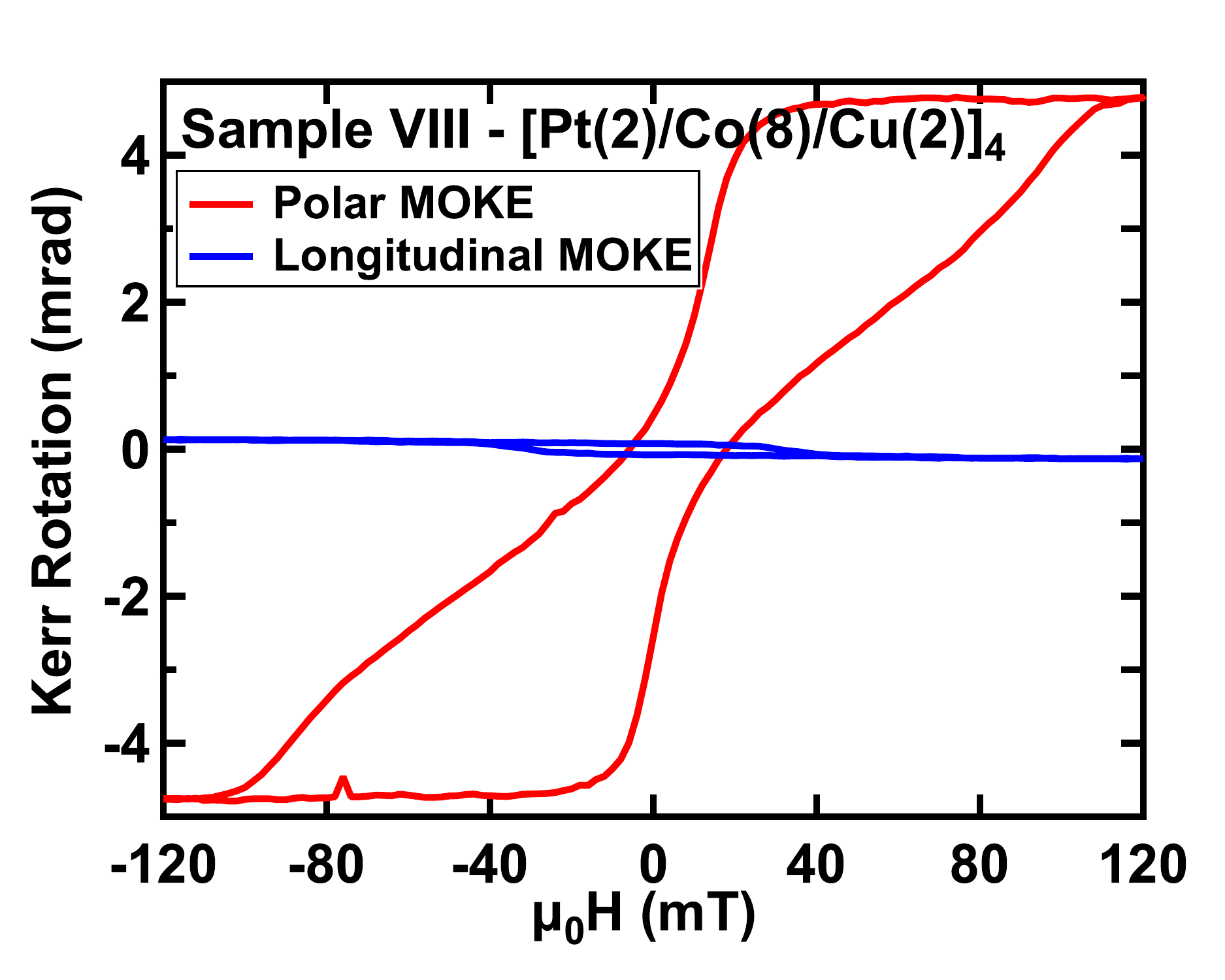}
    }
    \vfill
    \subfloat[\label{fig:MOKE_Structural}]{
       \includegraphics[width=\textwidth]{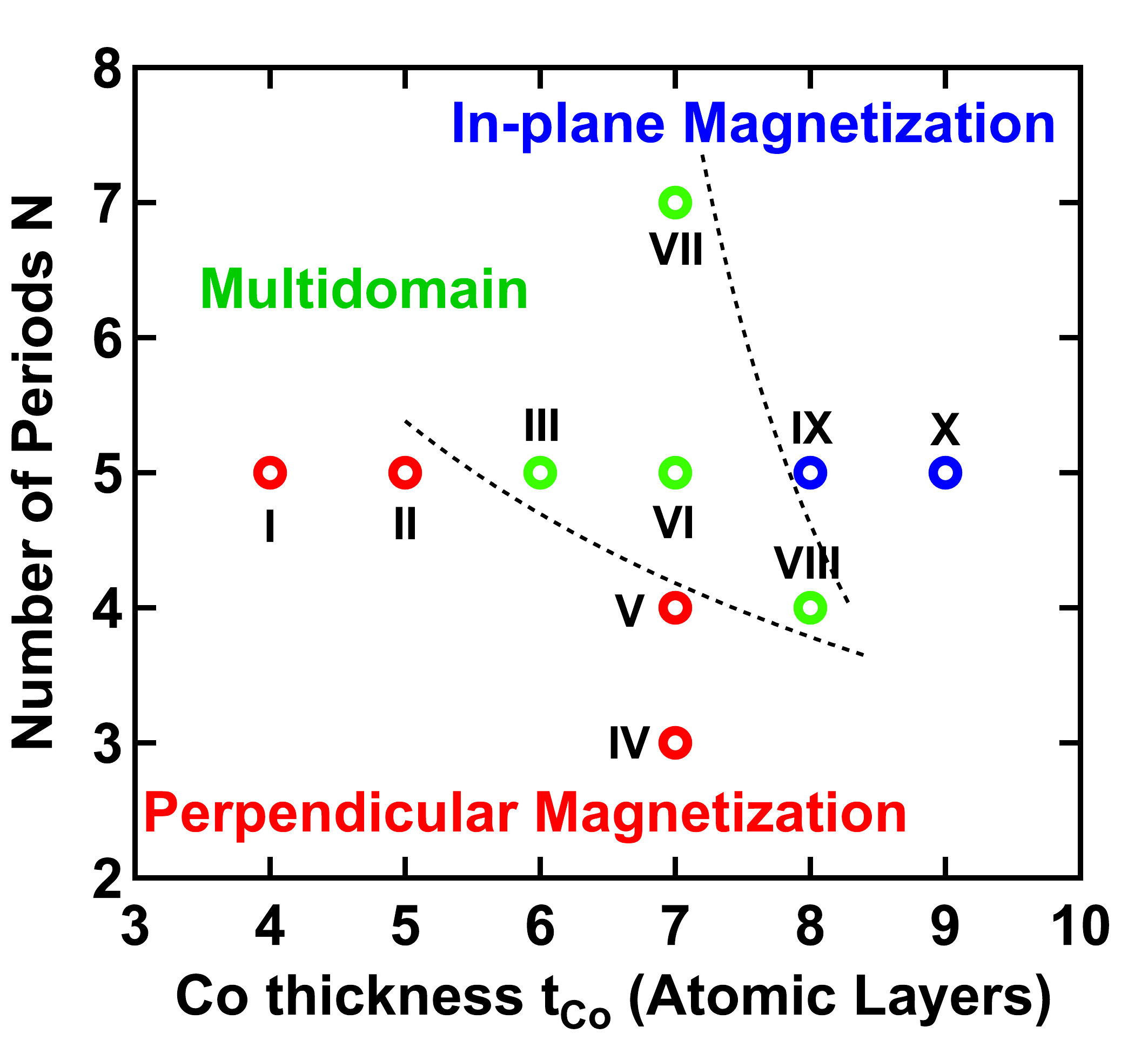}
    }
    \end{minipage}
    \caption{\label{fig:Structural} Relationship between MOKE hysteresis loops and the multilayer structure.
    (a) Representative polar (red curve) and longitudinal (blue curve) MOKE hysteresis loops of [Pt(2)/Co($t_{Co}$)/Cu(2)]$_5$ multilayers with a fixed number of periods $N$ = 5.
    (b) Representative polar (red curve) and longitudinal (blue curve) MOKE hysteresis loops of [Pt(2)/Co(7)/Cu(2)]$_N$ multilayers with fixed Co layer thickness $t_{Co}$ = 7\,AL.
    (c) Polar (red curve) and longitudinal (blue curve) MCD hysteresis loops of [Pt(2)/Co(7)/Cu(2)]$_7$ (sample VII) acquired on another setup.}
    (d) Polar (red curve) and longitudinal (blue curve) MOKE hysteresis loops of [Pt(2)/Co(8)/Cu(2)]$_4$ (sample VIII).
    (e) Summary of the sample series. The dashed lines are guides to the eye. 
\end{figure*}

\begin{figure*}[ht]
    \subfloat[\label{fig:LTEM_0mT}]{
        \includegraphics[width=0.33\textwidth]{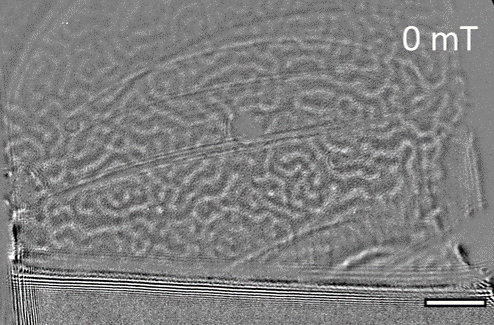}
    }
    \subfloat[\label{fig:LTEM_100mT}]{
        \includegraphics[width=0.33\textwidth]{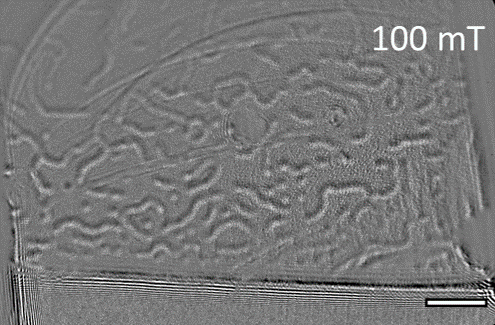}
    }
    \subfloat[\label{fig:LTEM_135mT}]{
        \includegraphics[width=0.33\textwidth]{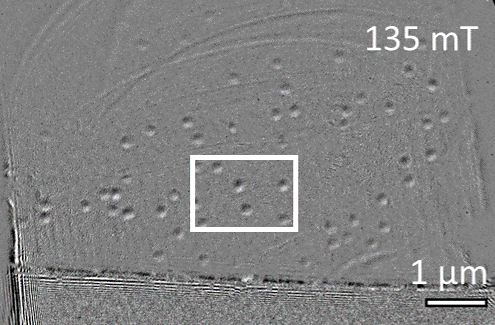}
    }\hfill
    \subfloat[\label{fig:LTEM_BGS}]{
        \includegraphics[width=0.33\textwidth]{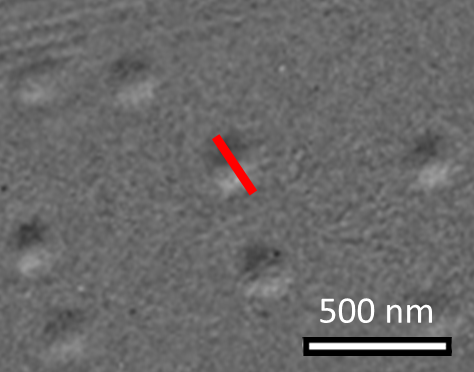}
    }
    \subfloat[\label{fig:LTEM_Linecut}]{
        \includegraphics[width=0.45\textwidth]{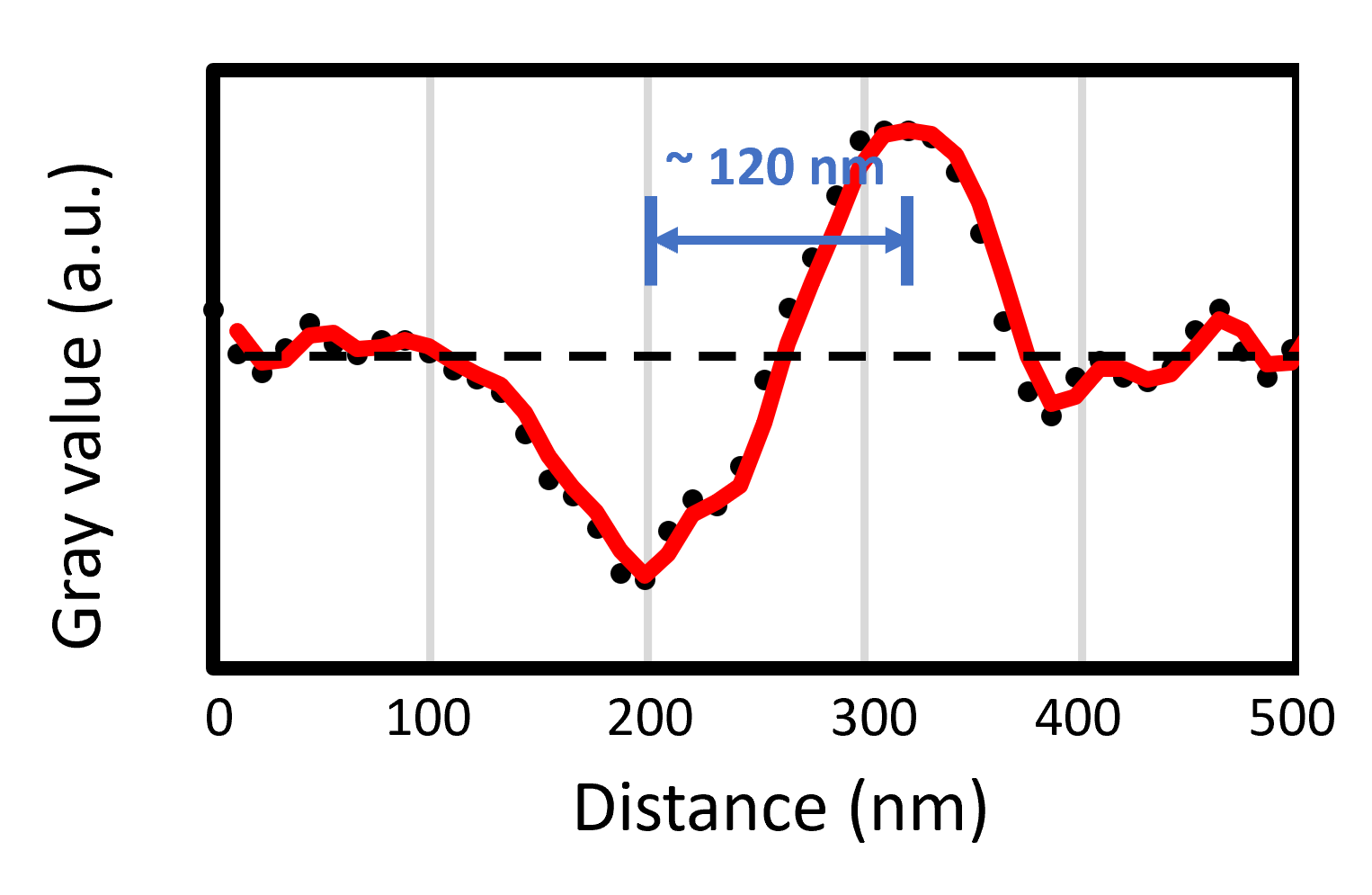}
    }
    \caption{\label{fig:LTEM} LTEM images of the [Pt(2)/Co(6)/Cu(2)]$_5$ thin film with a thinned substrate.
    (a-c) LTEM images of the [Pt(2)/Co(6)/Cu(2)]$_5$ sample under (a) 0\,mT, (b) 100\,mT, and (c) 135\,mT applied magnetic field. 
    The scale bar represents 1\,$\mu m$.
    (d). Zoomed-in LTEM image at 135 mT.
    The area of this image corresponds to the white box in Figure~\ref{fig:LTEM_135mT}.
    The scale bar represents 500\,nm. 
    (A background image taken in the field-polarized state (160\,mT) is subtracted from (a - d) to improve the contrast. 
    (e). Line-cut profile of a single skyrmion along the direction of the red line in Figure~\ref{fig:LTEM_BGS}.
    The red solid line is a smoothed curve of the line-cut profile.} 
\end{figure*}

\begin{figure*}[ht]
    \includegraphics[width=0.9\textwidth]{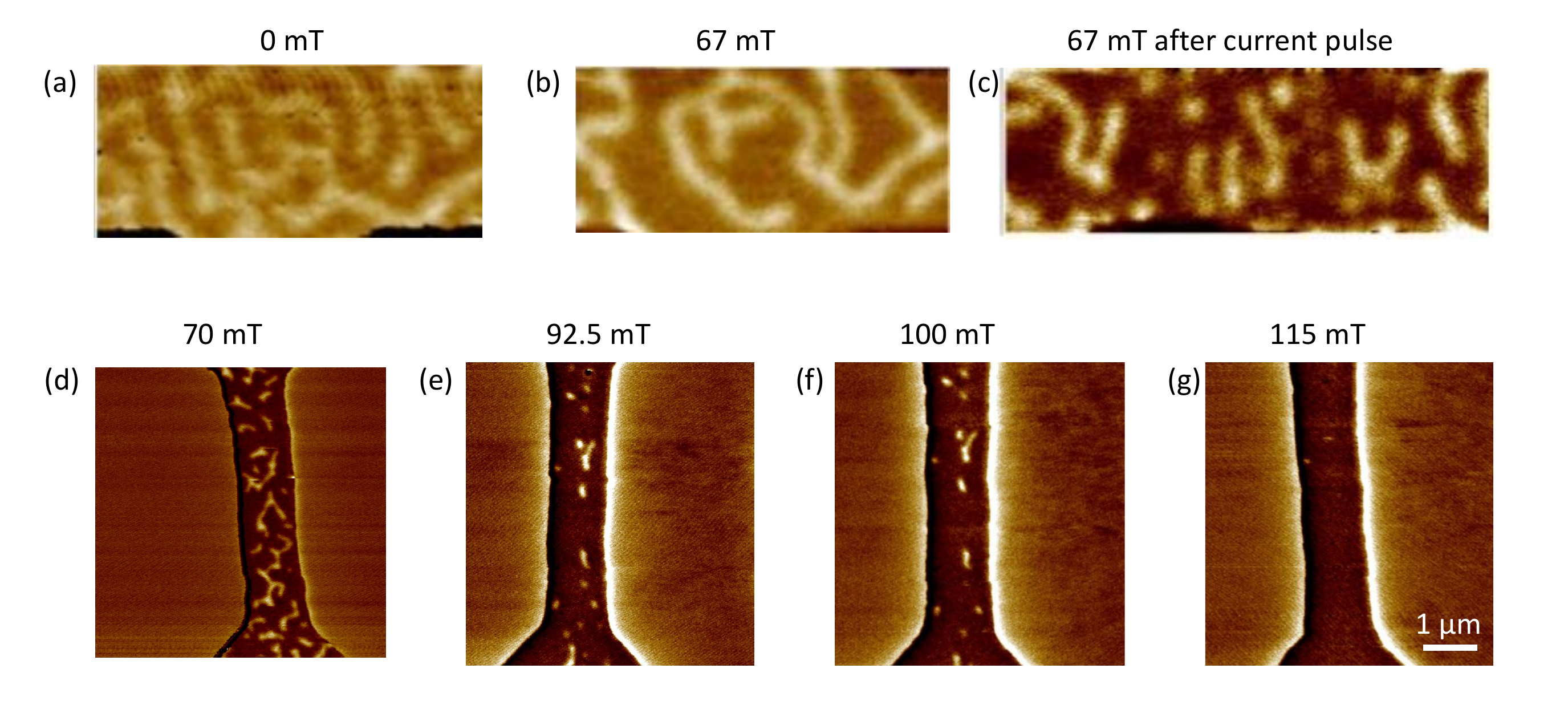}
    \caption{\label{fig:MFM_film} MFM images of skyrmions. 
    (a-c) For a [Pt(2)/Co(6)/Cu(2)]$_5$ sample, the magnetic field is ramped from -100\,mT to 0\,mT leading to labyrinth domains (a), ramping to 67\,mT produces domains with lower density (b), and application of a current pulse generates isolated skyrmions (c). 
    (d-f) For a [Pt(2)/Co(8)/Cu(2)]$_4$ sample, the magnetic field is ramped from -100\,mT to 70\,mT (d), 92.5\,mT (e), 100\,mT (f), and 115\,mT (g). With increasing field, magnetic textures evolve from stripe domains to skyrmions to a field-polarized state.
    } 
\end{figure*}

For tuning the hysteresis loop shape and magnetic anisotropy, we varied the Co thickness and the number of periods $N$. For these studies, we maintained a constant thickness of 2 AL for the Pt and Cu layers.

Beginning with the Co thickness, we synthesized a sample series of with $t_{Co}$ = 4, 5, 6, 7, 8 and 9 AL for a fixed number of periods $N = 5$. Representative polar (red curve) and longitudinal (blue curve) MOKE loops are shown in Figure~\ref{fig:MOKE_Structural_N5}. 
At a low Co thickness of 4\,AL (sample I), the polar loop is square with a large remanence while the longitudinal loop has a small signal with no remanence. 
This indicates a perpendicular (out-of-plane) magnetic easy axis.
As $t_{Co}$ increases, the polar MOKE hysteresis loop evolves to a wasp-waist shape with low remanence for $t_{Co}$ = 6\,AL (sample III), and eventually to almost linear with negligible hysteresis and zero remanences for $t_{Co}$ = 8\,AL
Meanwhile, the longitudinal MOKE hysteresis loop exhibits an increasing remanence with thickness, going from zero remanences for $t_{Co}$ = 4 and 6\,AL to a loop with substantial remanence and sharp magnetization reversals for 8\,AL.
This indicates a transition to in-plane magnetization for thicker Co. This spin reorientation transition from perpendicular magnetization to in-plane magnetization with increasing Co thickness is summarized by the horizontal points at $N = 5$ in Figure~\ref{fig:MOKE_Structural}, with red points signifying perpendicular magnetization, green points signifying the multidomain SRT region (the domain structure has been verified by MFM and LTEM elsewhere in this manuscript), and blue points signifying in-plane magnetization.

This thickness-dependent spin reorientation is understood as a competition between the perpendicular magnetic anisotropy (PMA) originating from the Pt/Co interface and the magnetic shape anisotropy favoring in-plane magnetization~\cite{zeper1989perpendicular}. The Pt/Co PMA is interfacial so its energy density scales inversely with Co thickness. On the other hand, the shape anisotropy is bulk-like so its energy density remains constant as Co thickness is varied. 
Thus, larger thicknesses will favor in-plane magnetization while smaller thicknesses will favor perpendicular magnetization.
We estimated the spin-reorientation thickness $d_{Co}$ of individual Co layer by balancing the uniaxial anisotropy and magnetic dipolar energy (see SM section S2 for details).
Taking magnetocrystalline anisotropy of hcp-structured cobalt $K_{MC}$ = 0.43$\times$10$^6$\,J/m$^3$ from~\cite{alameda1996effects}, and interfacial anisotropy $K_{S}$ = 0.9$\times$10$^{-3}$\,J/m$^2$ from the analysis of our SQUID data, the spin-reorientation thickness of cobalt is estimated to be 1.7\,nm.
This is in good agreement with our experimental results.

The hysteresis loops also exhibit variation as a function of the number of periods $N$ of the Pt/Co/Cu multilayer. Figure~\ref{fig:MOKE_Structural_7ALCo} shows the out-of-plane and in-plane hystersis loops for [Pt(2)/Co(7)/Cu(2)]$_N$ for N = 3, 5, and 7.
In this sample series, sample IV with 3 periods has a polar MOKE loop (red curve, top loop) with nearly 100\% remanence and a longitudinal MOKE loop (blue curve, top loop) with a weak response, indicating perpendicular magnetization. 
For sample VI with 5 periods, the polar MOKE loop (red curve, middle loop) has a wasp-waisted shape and the longitudinal MOKE loop shows a small signal with non-zero remanence. 
Similar to the wasp-waisted polar loops of Figure~\ref{fig:MOKE} and sample III in Figure~\ref{fig:MOKE_Structural_N5}, the loop is not fully saturated so these are minor loops. Nevertheless, the applied field goes sufficiently high to reveal the wasp-waisted loop corresponding to a multidomain state (see MFM in SM, Section S4). 
The easy axis is predominantly OOP, but the presence of hysteresis in the longitudinal loop (blue curve, middle loop) suggests the emergence of a small in-plane magnetization component.

For sample VII with 7 periods, we measured the hysteresis loops on a different magneto-optic setup with higher magnetic fields to saturate the magnetization fully. 
Here, we measured magnetic circular dichroism using a 532 nm laser at a 45$^\circ$ angle of incidence. 
As shown in the lower panel of Figure~\ref{fig:MOKE_Structural_7ALCo}, the polar MCD measurement (red curve) exhibits an elongated wasp-waisted loop that saturates at $\sim 170$ mT. 
The longitudinal MCD measurement (blue curve) shows the presence of an in-plane component of magnetization. 
These loops are characteristic of a multidomain state, which was directly confirmed by MFM (see SM, section S4). 
We note that the presence of multidomains for large $N$ is consistent with prior MOKE microscopy measurements of Pt/Co/Cu samples with $N = 10$~\cite{sun2016magnetic}. 
The transition from a perpendicularly magnetized state with high remanence at low $N$, to a multidomain structure with tilted hysteresis loops at higher $N$ has been known for Pt/Co~\cite{ochiai_copt_1989,lin_magnetic_1991,wang2020effect} and observed more recently in Pt/Co/X superlattices~\cite{wang2021manipulating}. 
This transition is likely due to stronger dipolar fields, and hence stronger dipolar couplings that drive the stabilization of domains, as $N$ increases. It is also worth noting that such dipolar fields could play an important role in stabilizing skyrmions~\cite{boulle_room-temperature_2016,buttner_theory_2018}.
This transition was also reproduced by our micromagnetic simulations, as discussed later.

To further investigate the boundary of the multidomain region, we synthesized a sample with $t_{Co}$ = 8\,AL and $N$ = 4 (sample VIII) that is to the lower-right of multidomain sample VI in the $N-t_{Co}$ diagram (Fig.~\ref{fig:MOKE_Structural}). The polar and longitudinal MOKE loops for this sample (Figure~\ref{fig:MOKE_20220516}) confirm the wasp-waisted polar loop that is characteristic of a multidomain state.

The zero-field magnetic states of the samples are summarized in Table~\ref{tab:table1} and the hysteresis loops of each sample are shown in Section 1 of the Supplementary Material (SM).

We now focus on the samples in the transition region with wasp-waisted polar hystersis loops, indicated by the green dots in Figure~\ref{fig:MOKE_Structural} (and labeled ``Multidomain'').
Here, we employed LTEM and MFM to image the magnetic domain structure and investigate the possible presence of skyrmions. 
For the LTEM measurements~\cite{wang2021stimulated}, focused ion beam (FIB) milling was utilized to thin the sapphire substrate, thus allowing the electron beam to transmit through the sample. 
Figure~\ref{fig:LTEM}a-c shows planar-view LTEM images of a [Pt(2)/Co(6)/Cu(2)]$_5$ multilayer sample at various out-of-plane magnetic fields. 
The straight parallel lines are from the substrate thinning and are not due to magnetic textures. 
For these images, we have tilted the sample by $20^\circ$ to achieve magnetic contrast for N\'{e}el skyrmions~\cite{benitez2015magnetic,mcvitie2018transmission}. 

Beginning at zero field (Figure~\ref{fig:LTEM_0mT}), we observe labyrinth magnetic domains which evolve into magnetic stripe domains when the field is ramped up to 100\,mT (Figure~\ref{fig:LTEM_100mT}).
When the field is raised to 135\,mT (Figure~\ref{fig:LTEM_135mT}), several magnetic bubbles are observed. 
The images in Figure~4(a - d) have improved contrast after subtracting a background image taken in the field-polarized state at 160\,mT~\cite{wang2022extracting}.

These bubbles are identified as N\'eel skyrmions based on the following considerations. 
First, there is no magnetic contrast when the electron beam is normally incident on the sample, which is the expected behavior since the Lorentz force is purely tangential for N\'{e}el skyrmions. 
Second, the magnetic contrast appears when the sample is tilted, and the bubble appears as having a positive and a negative lobe, which is the expected shape for N\'{e}el skyrmion~\cite{mcvitie2018transmission}. 
Section S3 of the SM provides a detailed explanation of the LTEM contrast, which depends on the sample tilt and not the skyrmion chirality.
A line-cut across the lobes, shown in Figure~\ref{fig:LTEM_Linecut} (see SM section S3 for line-cut profiles at different fields), establishes the size of the bubble to be approximately 120\,nm at 135\,mT. 

\begin{figure*}[ht]
  \begin{minipage}[b]{.4\textwidth}
    \subfloat[\label{fig:Simulation_0mT}]{
    \includegraphics[width=0.48\textwidth]{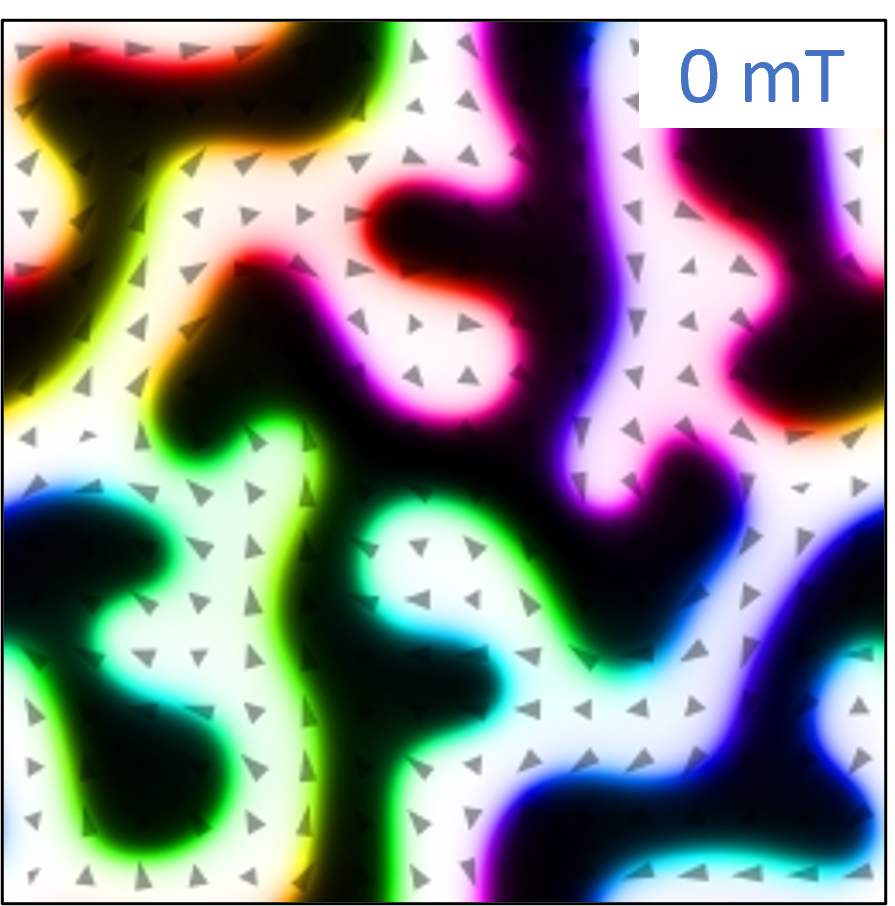}
    }
    \subfloat[\label{fig:Simulation_70mT}]{
    \includegraphics[width=0.48\textwidth]{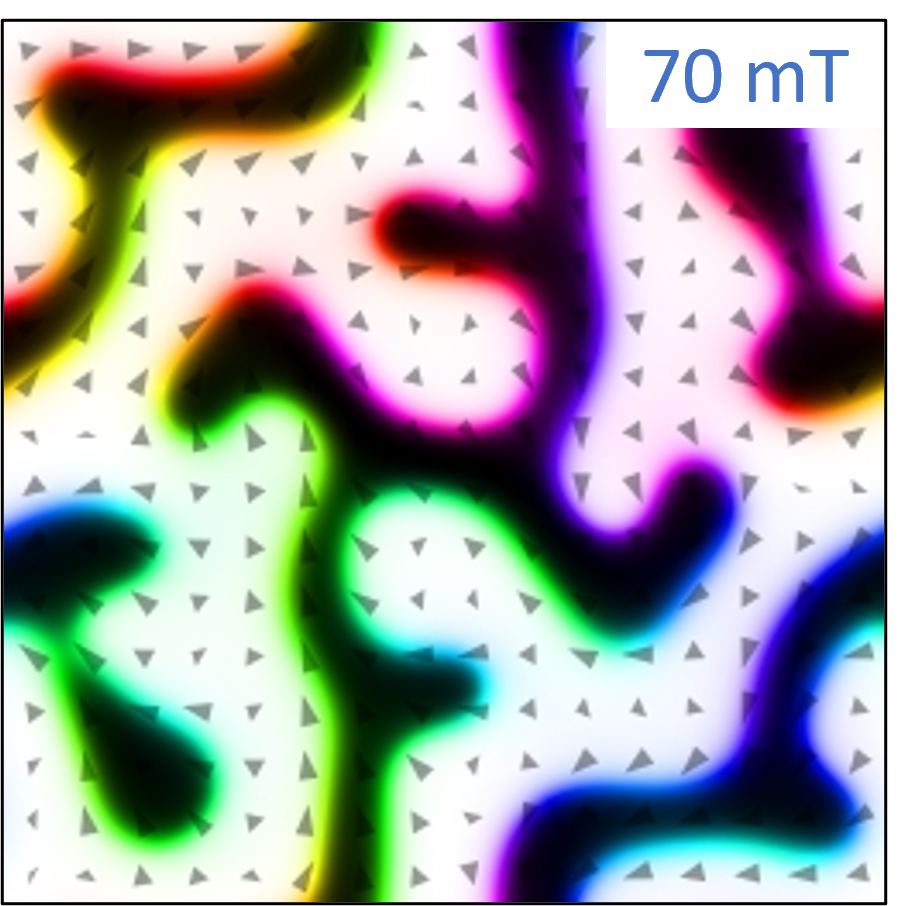}
    }\hfill
    \subfloat[\label{fig:Simulation_100mT}]{
    \includegraphics[width=0.48\textwidth]{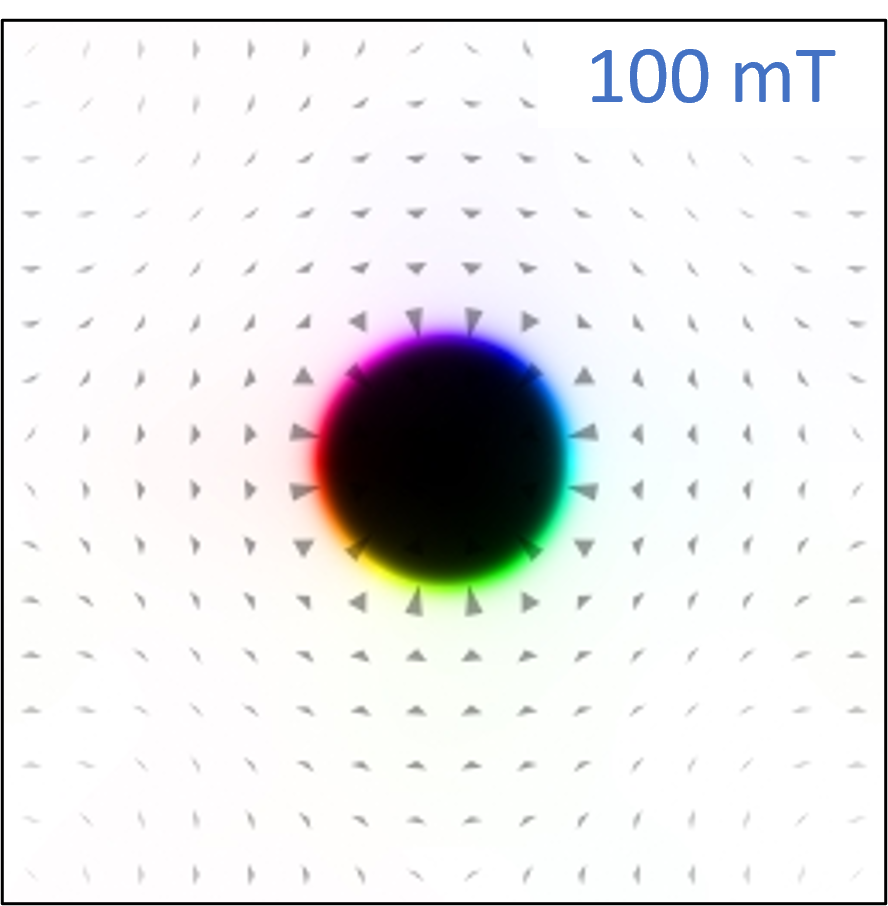}
    }
    \subfloat[\label{fig:Simulation_135mT}]{
    \includegraphics[width=0.48\textwidth]{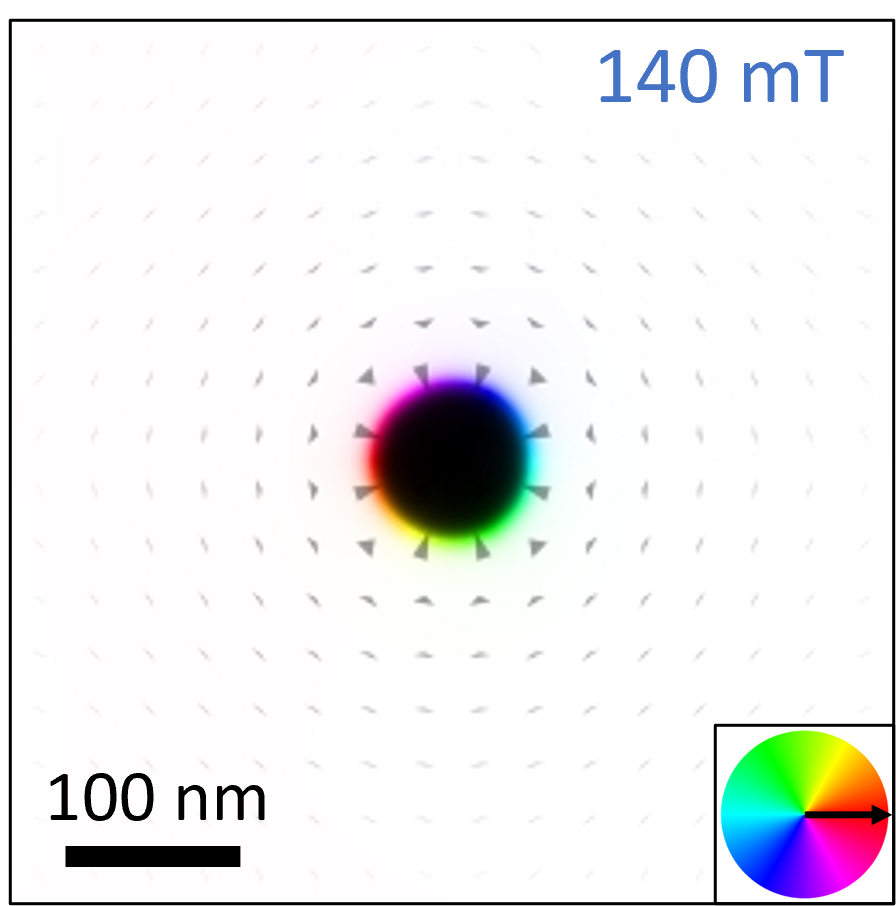}
    }
  \end{minipage}
  \begin{minipage}[b]{.5\textwidth}
    \subfloat[\label{fig:Simulation_size}]{
    \includegraphics[width=\textwidth]{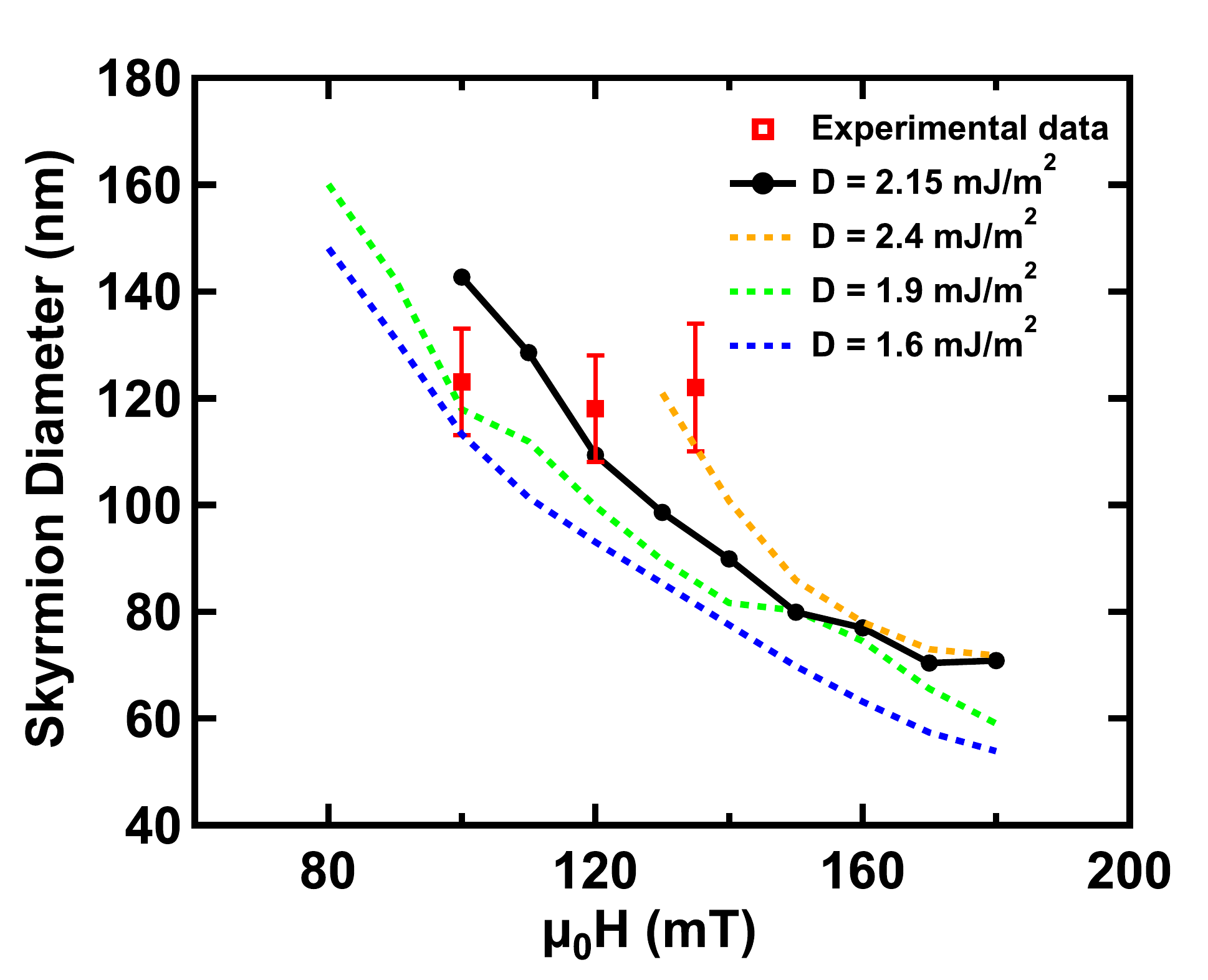}
    }
  \end{minipage} 
    \caption{\label{fig:simulation} (a-d) Micromagnetic simulation of [Pt(2)/Co(6)/Cu(2)]$_5$ sample with (a) 0\,mT, (b) 70\,mT, (c) 100\,mT (d) 140\,mT applied field. The color wheel represents the in-plane component of the magnetization direction. (e). Simulated diameter of skyrmions as a function of applied field using several different DMI values. 
    The red stars represent experimental data from LTEM measurements (see SM section S3 for line-cut profiles at different fields)}.
\end{figure*}

We further investigate the magnetic domain structure and skyrmion spin textures using MFM. 
An advantage of MFM over LTEM is that sample preparation including substrate thinning is not required. 
This provides access to the as-grown magnetic properties, as substrate thinning could introduce strain. 
In addition, MFM measurements are compatible with devices fabricated by photolithography and electron-beam lithography. 
The measurements were performed in the Bruker MFM equipped with a homemade variable magnet with out-of-plane field range of -115\,mT to 115\,mT. 
The [Pt/Co/Cu] multilayers were patterned to have micron-wide device channels.

Upon investigating samples in the ``Multidomain'' region of the phase diagram (green dots in Figure~\ref{fig:MOKE_Structural}), we found that skyrmions could be nucleated either using field ramp sequences or current pulses. 
For a [Pt(2)/Co(6)/Cu(2)]$_5$ sample (Figure~\ref{fig:MFM_film}a-c), the out-of-plane magnetic field was first set to -100\,mT. 
Ramping to 0\,mT produced labyrinth magnetic domains in the channel (Figure~\ref{fig:MFM_film}a).
Fast Fourier transform (FFT) analysis of the MFM image yields an average labyrinth domain width of 130\,nm at zero field.
Increasing the field to 67\,mT, the domain structure evolved to magnetic stripe domains with a lower density of domain walls (Figure~\ref{fig:MFM_film}b), consistent with the domain structures observed by LTEM (Figure~\ref{fig:LTEM_100mT}). 
As the MFM magnet is unable to reach the 135\,mT needed to nucleate skyrmions, we decided to use current pulses to help nucleate skyrmions, as demonstrated previously in other materials~\cite{jiang2015, woo2016observation, lemesh_current-induced_2018, juge2022skyrmions}. 
Following a single current pulse of 1.39$\times10^{12}$\,A/m$^2$ with duration 20\,ns, some of the stripe domains have broken up into isolated skyrmions (Figure~\ref{fig:MFM_film}c). 

For a [Pt(2)/Co(8)/Cu(2)]$_4$ sample, the saturation field is within the range of the MFM magnet, so we investigate skyrmion nucleation by field ramping. 
Starting by applying a -100\,mT field, we ramp up through 0\,mT to a final field of 115\,mT. Figure~\ref{fig:MFM_film}d-g show MFM images at representative fields. 
At 70\,mT, the magnetic texture is dominated by stripe domains and a few skyrmions are observed. 
Increasing to 92.5\,mT causes many of the stripes to nucleate into skyrmions or disappear altogether. 
At 100\,mT, some of the skyrmions have disappeared, along with the stripes. 
Finally, at 115\,mT, most of the sample has become field-polarized. 

Additional MFM measurements establish the presence of skyrmions in a [Pt(2)/Co(7)/Cu(2)]$_5$ sample (SM Section S4).

Micromagnetic simulations using MuMax3~\cite{vansteenkiste2014design} were performed to try to understand two important aspects of [Pt/Co/Cu]$_N$ multilayers - out-of-plane ferromagnetic domain transition to multidomain transition, as well as the size of skyrmions.
In our simulation, the [Pt(2)/Co(6)/Cu(2)]$_N$ was modeled as a $3N$-layer stack, with the magnetic moment of the Co and Pt layers, the exchange coupling between the neighboring Co and Pt layers, and interlayer exchange coupling between neighboring Co layers taken into consideration.
The interlayer exchange coupling between Co layers through Pt/Cu was set to be $\sim$10\% of the exchange coupling within Co layers~\cite{mukhopadhyay2020asymmetric}.
The micromagnetic parameters used were exchange stiffness $A_{ex}$ of 25\,pJ/m~\cite{grimsditch_exchange_1997, ding_magnetic_2005, legrand_spatial_2022}, uniaxial anisotropy $K_u$ of 1.2\,MJ/m$^3$, DMI strengths $D$ of 2.15\,mJ/m$^2$ from \cite{jia2020material}, and saturation magnetization $M_S$ of 1.4$\times$10$^6$\,A/m and 1.7$\times$10$^6$\,A/m for Co and Pt layers, respectively.
The selection of each micromagnetic parameter is justified in the SM section S5.

We first investigated the transition from uniformly magnetized states to multidomain states as the number of periods $N$ increases.
In these simulations, the initial state was magnetized along the out-of-plane direction, with 1\% of spins randomized to avoid unstable equilibrium states.
Then the system was allowed to relax from the initial state at zero field.
For $N$ = 1 to 4, the magnetization remains uniformly magnetized at zero field, while for $N$ = 5, the uniformly magnetized state breaks into labyrinth domains at zero field (see SM Section S5 for details).
This is because as $N$ increases, the dipolar interaction increases and overcomes the exchange interaction that favors a uniformly magnetized state~\cite{boulle_room-temperature_2016, buttner_theory_2018, lemesh_current-induced_2018}.
FFT analysis yields a labyrinth domain width of 120\,nm at zero fields, in good agreement with the MFM data.

Many different values of A$_{ex}$ were found in the literature, ranging from 10\,pJ/m to 35\,pJ/m~\cite{skomski_appendix_2008, ding_ultra_2002, vernon1984brillouin, grimsditch_exchange_1997, ding_magnetic_2005, legrand_spatial_2022, liu1996exchange, shirane1968spin}. 
To investigate the influence of A$_{ex}$ on multidomain formation, we lowered $A_{ex}$ from 25\,pJ/m to 10\,pJ/m, which is another widely used value~\cite{skomski_appendix_2008}.
With smaller exchange stiffness, all of [Pt(2)/Co(6)/Cu(2)]$_N$ multilayers relaxed into multidomain states, because such a small exchange stiffness is not sufficient to maintain a uniformly magnetized state (see SM section S5 for details).
In addition, with A$_{ex}$ = 10\,pJ/m, the simulated labyrinth domain width is 55\,nm, much smaller than the experimentally observed value.
Therefore, the agreement of simulation and experiment for 25\,pJ/m but not for 10\,pJ/m indicates the exchange stiffness of cobalt in our samples is close to 25\,pJ/m.

We next studied the skyrmion size and its dependence on various parameters using the same micromagnetic model.
In these simulations, the system relaxes from a N{\'e}el skyrmion with $\sim$75\,nm diameter.
Results of the simulations for the [Pt(2)/Co(6)/Cu(2)]$_5$ sample are shown in Fig.~\ref{fig:simulation}. 
At zero field, the simulations show a labyrinth domain structure, consistent with both the MFM and LTEM (Fig.~\ref{fig:Simulation_0mT}). 
As the magnetic field is increased, N{\'e}el skyrmions can be stabilized starting around 80-90\,mT.
Continuing to increase the field causes the skyrmions to shrink, as shown in Fig.~\ref{fig:Simulation_100mT} and~\ref{fig:Simulation_135mT}.
Comparing with experimental values of skyrmion size that show little variation with field (red squares in Fig.~\ref{fig:Simulation_size}), the simulated skyrmion size exhibits a stronger variation with field than is consistent with the experimental uncertainties (see SM section S3 for details on the analysis). 
The reason for this discrepancy is unclear. Nevertheless, the experimental values of the skyrmion size near 120\,nm
suggest that the value of the DMI falls in between 1.9\,mJ/m$^2$ and 2.4\,mJ/m$^2$ (see Fig.~\ref{fig:Simulation_size}). 
Another trend observed in the simulation is an increase in the skyrmion size with increasing values of $D$. 
These simulated dependencies of skyrmion size on $D$ and magnetic field are in agreement with previous analytical results in Wang \emph{et al.}~\cite{wang2018theory}.

The variation of skyrmion size with $D$ is understood as follows.
Intuitively, a skyrmion can be thought as being composed of an inner ``core'' region with magnetization antiparallel to the applied field, and an intermediate ``wall'' region where the magnetization rotates until it becomes parallel to the applied field in the ``outside'' region. 
The Zeeman energy is most important in the core and outside regions, so increasing the field will shrink the core and increase the outside ferromagnetic region to lower the Zeeman energy, which leads to a smaller skyrmion. 
The DMI is most important of the wall region, so with a larger $D$, the skyrmion can lower its overall energy by increasing the width of the wall even though the Zeeman and anisotropy energies will increase. 
Thus, the skyrmion size will get larger with increasing $D$.

In conclusion, we observed room-temperature skyrmions in [Pt/Co/Cu]$_N$ multilayers. 
By varying the number of periods $N$ and the Co thickness, we tuned the magnetic state from perpendicularly magnetized to multidomain (wasp-waisted loop shape) to in-plane magnetized. 
Magnetic imaging by LTEM and MFM on multidomain samples showed that the magnetic spin texture evolves from labyrinth domains at low perpendicular magnetic fields to isolated skyrmion spin textures at higher fields. 
By tilting the sample during LTEM, we verified that the skyrmions are N\'eel type.  
MFM measurements on patterned devices showed that current pulses could nucleate skyrmions at lower magnetic fields compared to only ramping magnetic fields. 
While our imaging studies focused on the multidomain samples, we do not exclude the possibility of skyrmions in the OOP- or IP-magnetized samples.
This work establishes Pt/Co/Cu multilayers as a model system for investigating room-temperature skyrmions in ultrathin epitaxial materials.

\section*{Acknowledgments}
We acknowledge stimulating discussions with Denis Pelekhov.
This work was supported by the DARPA TEE program under Grant No. D18AP00008. Z.L. was supported by AFOSR/MURI project 2DMagic FA9550-19-1-0390 and US Department of Energy DE-SC0016379.
This research was partially supported by the Center for Emergent Materials, an NSF MRSEC, under award number DMR-2011876. 
Electron microscopy was performed at the Center for Electron Microscopy and Analysis (CEMAS) at The Ohio State University.

\section*{Author contributions}

S.C. synthesized the materials and performed the RHEED, MOKE, and SQUID measurements. 
N.B. and B.W. performed the TEM and LTEM measurements. 
C.M.S., Z.L., and S.D. performed the MFM measurements.
Z.L. performed MCD measurements.
S.C., J.B.F., M.R. and R.K.K. performed micromagnetic simulations.
R.K.K., D.W.M, M.R., and P.C.H. conceived the study. 
All authors participated in data analysis and preparation of the manuscript.

\bibliography{PtCoCu.bib}

\end{document}



\title{Supplemental Material: Room-Temperature Magnetic Skyrmions in Pt/Co/Cu Multilayers}

\author{Shuyu Cheng}
\affiliation{Department of Physics, The Ohio State University, Columbus, Ohio 43210, United States}
\author{N\'uria Bagu\'es}
\affiliation{Department of Materials Science and Engineering, The Ohio State University, Columbus, Ohio 43210, United States}
\author{Camelia M. Selcu}
\email{selcu.1@osu.edu}
\affiliation{Department of Physics, The Ohio State University, Columbus, Ohio 43210, United States}
\author{Jacob B. Freyermuth}
\affiliation{Department of Physics, The Ohio State University, Columbus, Ohio 43210, United States}
\author{Ziling Li}
\affiliation{Department of Physics, The Ohio State University, Columbus, Ohio 43210, United States}
\author{Binbin Wang}
\affiliation{Department of Materials Science and Engineering, The Ohio State University, Columbus, Ohio 43210, United States}
\author{Shekhar Das}
\affiliation{Department of Physics, The Ohio State University, Columbus, Ohio 43210, United States}
\author{P. Chris Hammel}
\affiliation{Department of Physics, The Ohio State University, Columbus, Ohio 43210, United States}
\author{Mohit Randeria}
\affiliation{Department of Physics, The Ohio State University, Columbus, Ohio 43210, United States}
\author{David W. McComb}
\email{mccomb.29@osu.edu}
\affiliation{Department of Materials Science and Engineering, The Ohio State University, Columbus, Ohio 43210, United States}
\author{Roland K. Kawakami}
\email{kawakami.15@osu.edu}
\affiliation{Department of Physics, The Ohio State University, Columbus, Ohio 43210, United States}

\maketitle

\newpage

\tableofcontents

\newpage

\section{\NoCaseChange{MOKE hysteresis loops of the [Pt/Co/Cu] multilayers}}

\begin{table*}[ht]
\begin{tabular*}{\textwidth}{c @{\extracolsep{\fill}} cccc}
 Sample ID & Co thickness t$_{Co}$ & Number of periods N & Zero-field magnetic state \\\hline
 I & 4 & 5 & OOP \\
 II & 5 & 5 & OOP \\
 III & 6 & 5 & Multidomain \\
 IV & 7 & 3 & OOP \\
 V & 7 & 4 & OOP \\
 VI & 7 & 5 & Multidomain \\
 VII & 7 & 7 & Multidomain \\
 VIII & 8 & 4 & Multidomain \\
 IX & 8 & 5 & IP \\
 X & 9 & 5 & IP
\end{tabular*}
\caption{\label{tab:tableS1} Summary of the structures and magnetic properties of the samples. This is a replica of Table.~1 in the main text.}
\end{table*}

\begin{figure*}
    \includegraphics[width=0.8\textwidth]{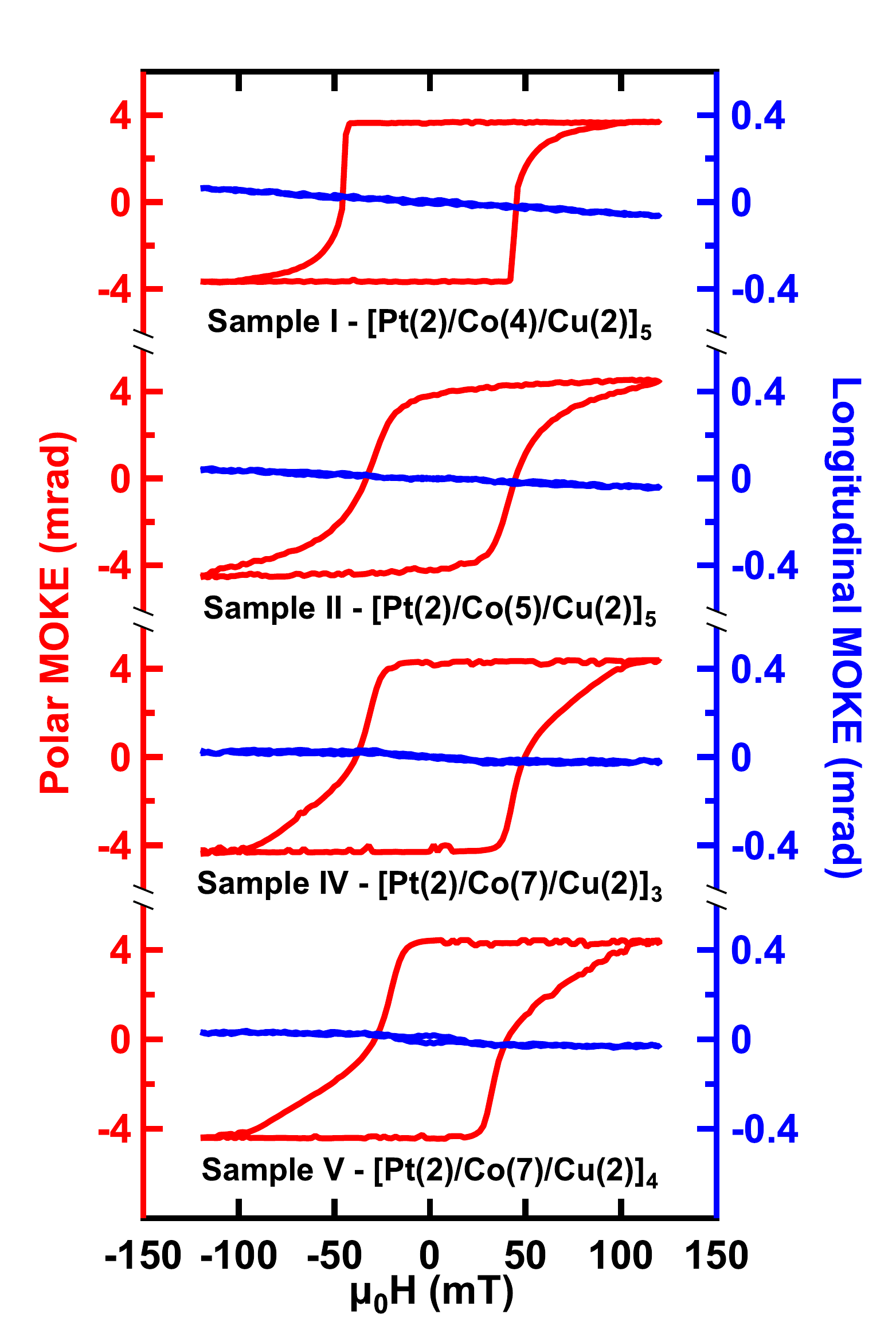}
    \caption{\label{fig:MOKE_Series_OOP}Polar (red) and longitudinal (blue) MOKE hysteresis loops of [Pt/Co/Cu] multilayers with strong easy-axis anisotropy.
    } 
\end{figure*}

\begin{figure*}
    \includegraphics[width=0.8\textwidth]{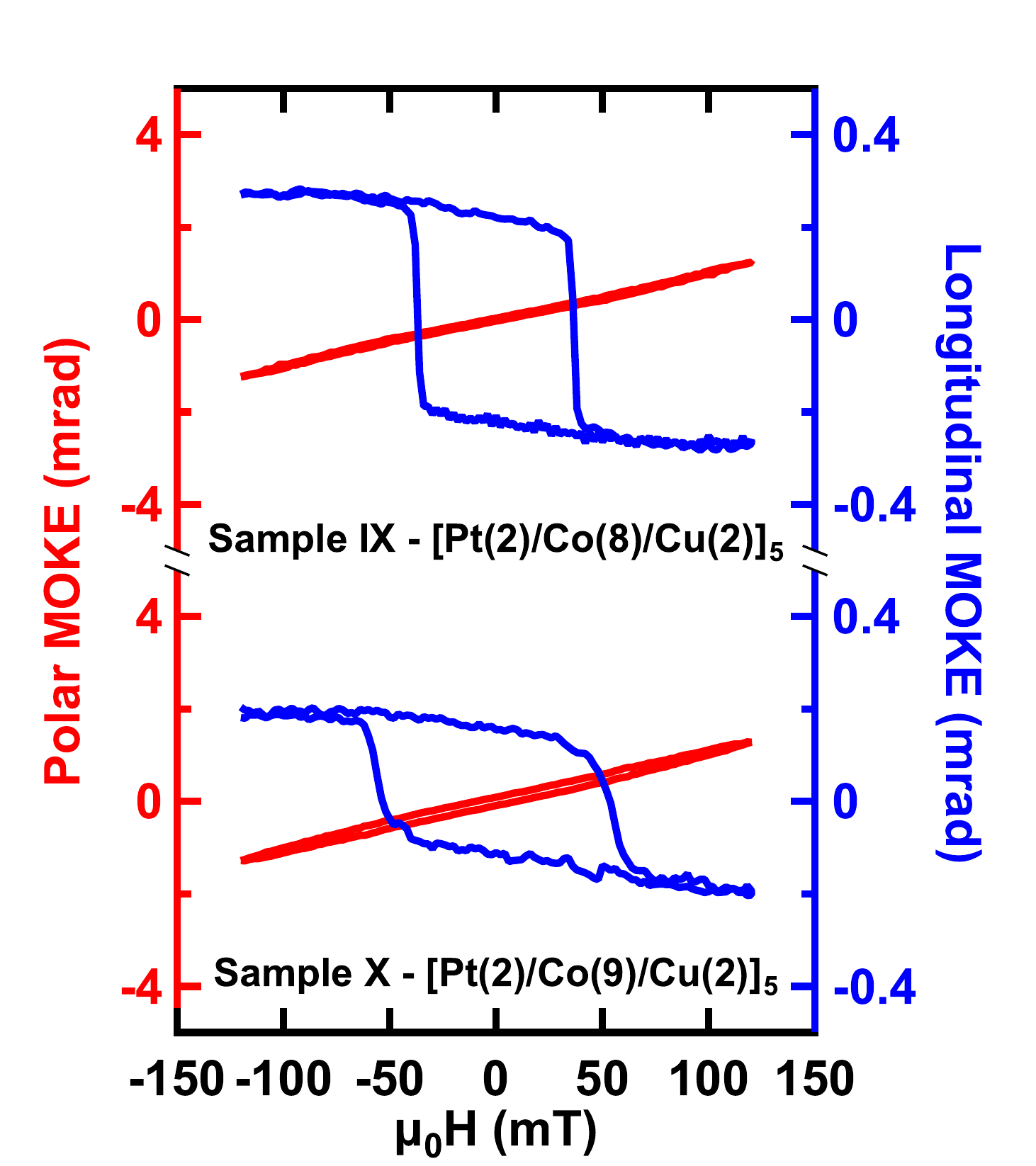}
    \caption{\label{fig:MOKE_Series_IP}Polar (red) and longitudinal (blue) MOKE hysteresis loops of [Pt/Co/Cu] multilayers with easy-plane anisotropy.
    } 
\end{figure*}

\begin{figure*}
    \includegraphics[width=0.8\textwidth]{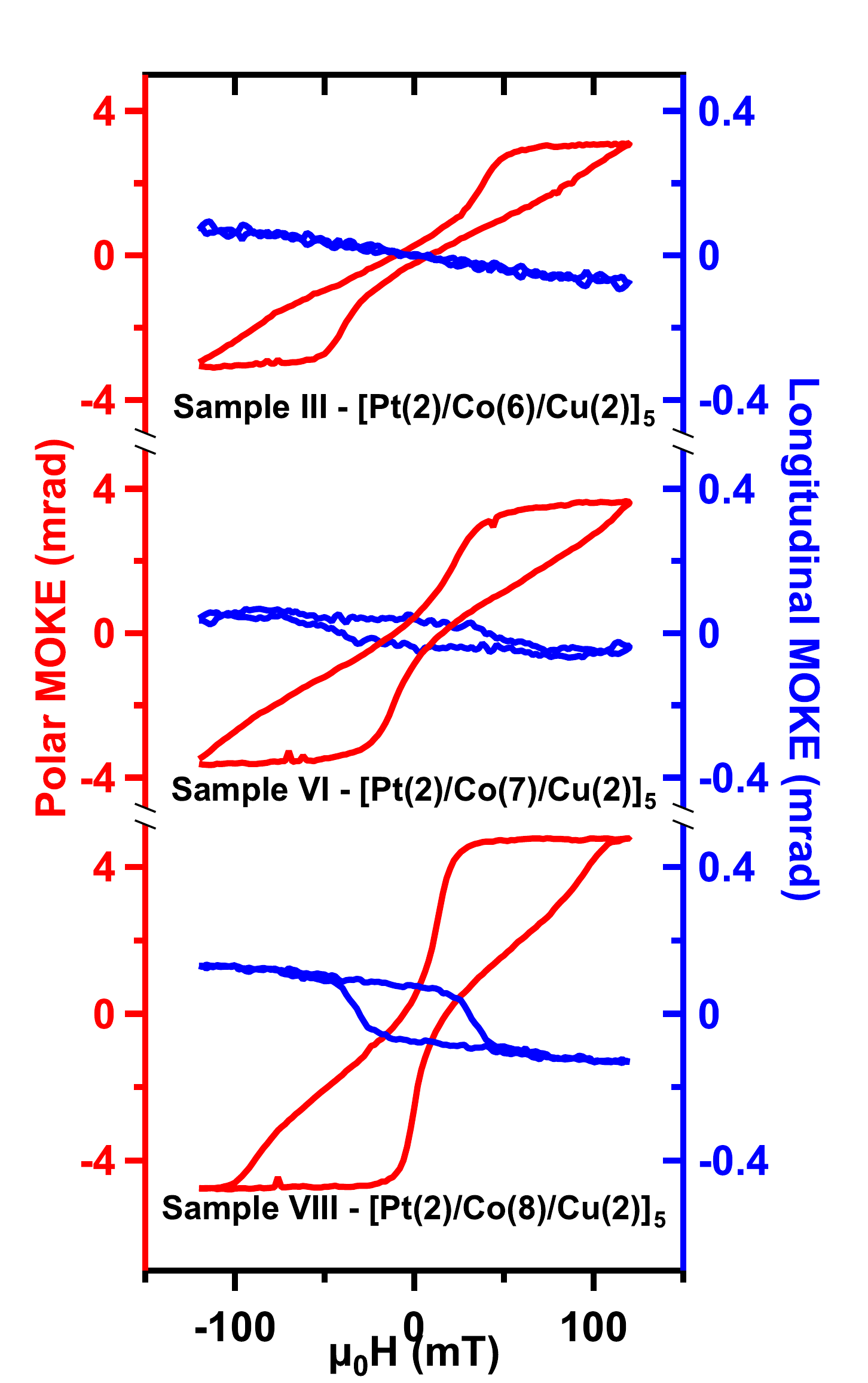}
    \caption{\label{fig:MOKE_Series_MD}Polar (red) and longitudinal (blue) MOKE hysteresis loops of [Pt/Co/Cu] multilayers with multidomain states at zero field.
    } 
\end{figure*}

\begin{figure*}
    \includegraphics[width=0.8\textwidth]{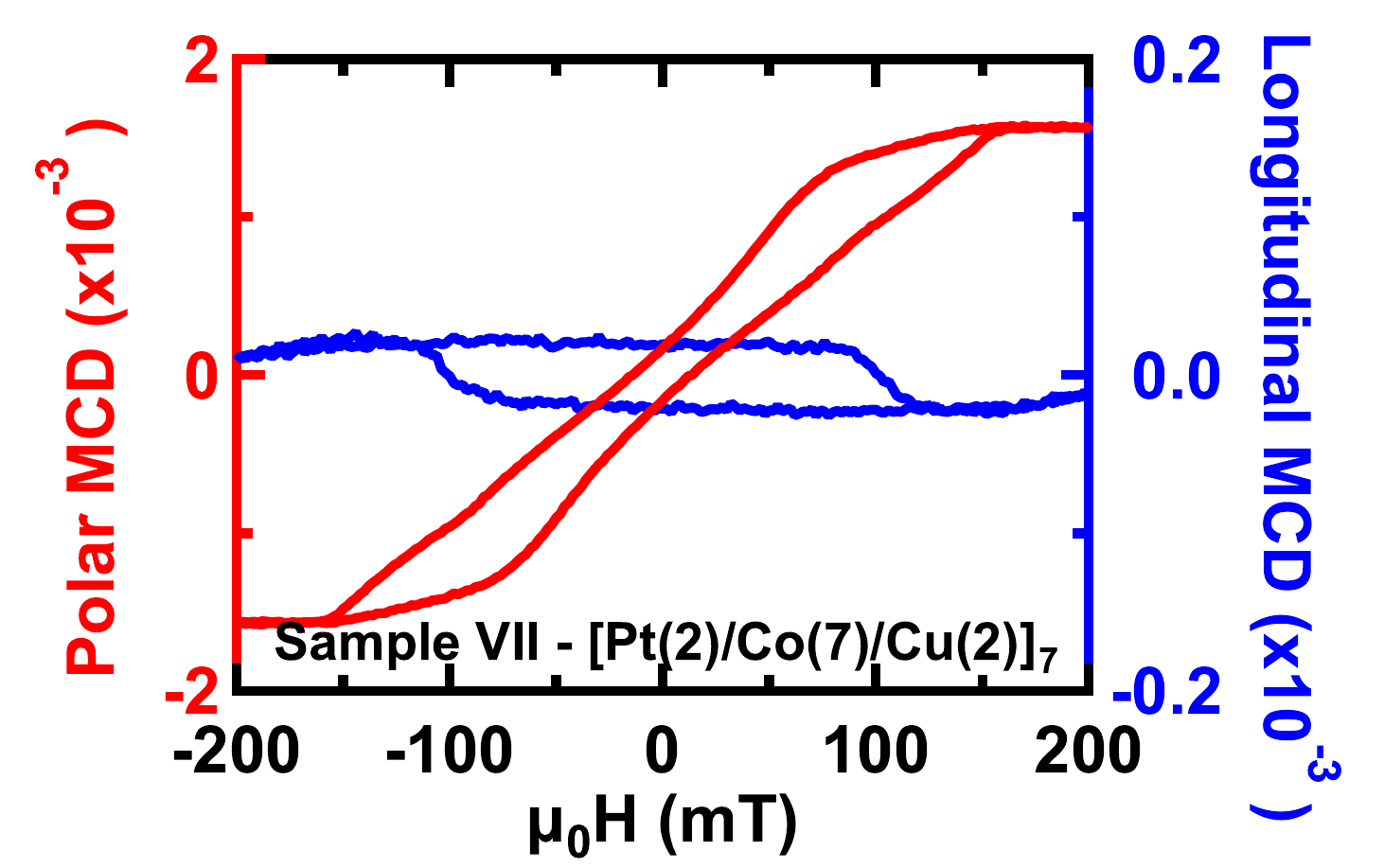}
    \caption{\label{fig:MOKE_Series_MD}Polar (red) and longitudinal (blue) MCD hysteresis loops of [Pt/Co/Cu] multilayers with multidomain states at zero field.
    } 
\end{figure*}

The structures and magnetic properties of [Pt(2)/Co(t$_{Co}$)/Cu(2)]$_N$ sample series are summarized in Table.~\ref{tab:tableS1} (replica of Table.~1 in the main text).
Fig.~\ref{fig:MOKE_Series_OOP} shows the MOKE hysteresis loops of [Pt(2)/Co(t$_{Co}$)/Cu(2)]$_N$ samples with strong easy-axis anisotropy (sample I, II, IV, V).
Fig.~\ref{fig:MOKE_Series_IP} shows the MOKE hysteresis loops of [Pt(2)/Co(t$_{Co}$)/Cu(2)]$_N$ samples with easy-plane anisotropy (sample VII, IX, X).
Fig.~\ref{fig:MOKE_Series_MD} shows the MOKE hysteresis loops of [Pt(2)/Co(t$_{Co}$)/Cu(2)]$_N$ samples with multidomain states at zero field (sample III, VI, VIII).

\clearpage

\section{\NoCaseChange{SQUID measurements of the [Pt/Co/Cu] multilayers}}

The SQUID measurements were performed using VSM mode in a Quantum Design SQUID magnetometer in both in-plane and out-of-plane geometries.
A linear diamagnetic background was subtracted from the raw magnetic moment data, and the magnetization was obtained by normalizing the magnetic moment by the volume of the sample.
The lateral size of the sample was measured by the caliper, while the thickness of the sample was given by the growth rate multiplied by the growth time of each layer.

For a typical 5\,mm$\times$5\,mm sample, the magnetic moment is $\sim$4$\times$10$^{-4}$\,emu, while the standard error from VSM measurement is usually less than $\sim$3$\times$10$^{-7}$\,emu.
Therefore, the uncertainty of the magnetic moment is less than 0.1\%.
The caliper measurement of sample size has an uncertainty of 0.03\,mm, leading to an uncertainty of 0.1\% in lateral size measurements.
The dominant uncertainty is the uncertainty in the film thickness measurement which is $\sim$10\%. 

We determined the interfacial anisotropy from SQUID data.
The total effective anisotropy was estimated from the difference between IP and OOP saturation fields:
\begin{equation}
   K_{eff} = \frac{1}{2}(\mu_0 H_{S}^{IP}-\mu_0 H_{S}^{OOP})M_S
\end{equation}
For [Pt(2)/Co(6)/Cu(2)]$_5$, by substituting in IP saturation field $\mu_0 H_{S}^{IP}$= 0.70\,T, OOP saturation field $\mu_0 H_{S}^{OOP}$= 0.12\,T, and an averaged saturation magnetization $M_S = (M_{Co}t_{Co}^{total}+M_{Pt}t_{Pt}^{total})/(t_{Co}^{total}+t_{Pt}^{total}+t_{Cu}^{total})$ = 1.18$\times$10$^3$\,kA/m, we got effective anisotropy $K_{eff}$ = 0.34$\times$10$^6$\,J/m$^3$.
Here we used $t_{Pt}^{total}$ = 2.3\,nm, which is the thickness of 10 atomic layers, because most of the Pt moment concentrate in the first two atomic layers of Pt adjacent to the Co layers~\cite{mukhopadhyay2020asymmetric}.
The total uniaxial anisotropy $K_{u}$ is given by:
\begin{equation}
   K_{u} = K_{eff} + \frac{1}{2}\mu_0 M_S^2
\end{equation}
By substituting in $M_S$ = 1.18$\times$ 10$^3$\,kA/m, we got $K_{u}$ = 1.2$\times$10$^6$\,J/m$^3$.
In [Pt/Co/Cu]$_N$ multilayers, the major contributions to the uniaxial anisotropy are magnetocrystalline anisotropy and interfacial anisotropy.
Taking the magnetocrystalline anisotropy of hcp-structured cobalt $K_{MC}$ = 0.43$\times$10$^6$\,J/m$^3$ from~\cite{alameda1996effects}, and the thickness of a single Co layer $t_{Co}$ = 1.2\,nm, we found the interfacial anisotropy of a Pt/Co interface to be $K_S=(K_{u}-K_{MC})t_{Co}$ = 0.9$\times$10$^{-3}$\,J/m$^2$ (i.e.~favoring perpendicular magnetization).
This value for interfacial anisotropy is consistent with the results from magnetic torque measurements on MBE-grown and sputtered Pt(111)/Co multilayers (0.97$\times$10$^{-3}$\,J/m$^2$ and  0.92$\times$10$^{-3}$\,J/m$^2$, respectively)~\cite{weiler1993interface}.

The spin-reorientation thickness of cobalt can be estimated by balancing the uniaxial anisotropy and the magnetic dipolar (shape) anisotropy:
\begin{equation}\label{eq:SRT}
   K_{eff} = \frac{K_{S}}{t_{Co}} + K_{MC} - \frac{1}{2}\mu_0 M_S^2 = \frac{K_{S}}{t_{Co}} + K_{MC} - \frac{1}{2}\mu_0\left(\frac{M_{Co}t_{Co}+M_{Pt}t_{Pt}}{t_{Co}+t_{Pt}+t_{Cu}}\right)^2 = 0
\end{equation}
Here we assume the [Pt/Co/Cu]$_N$ multilayers have similar induced Pt magnetization $M_{Pt}$ = 1.56$\times$10$^3$\,kA/m for larger $t_{Co}$.
Other parameters used are $M_{Co}$ = 1.43$\times$10$^3$\,kA/m, $t_{Pt}$ = 0.45\,nm, and $t_{Cu}$ = 0.41\,nm.
Numerically solving Equation~\ref{eq:SRT} gives spin re-orientation thickness of cobalt $t_{Co}$ = 1.7\,nm, which is in good agreement with our experimental results.

\begin{figure*}[ht]

    \subfloat[\label{fig:SQUID_722}]{
    \includegraphics[width=0.48\textwidth]{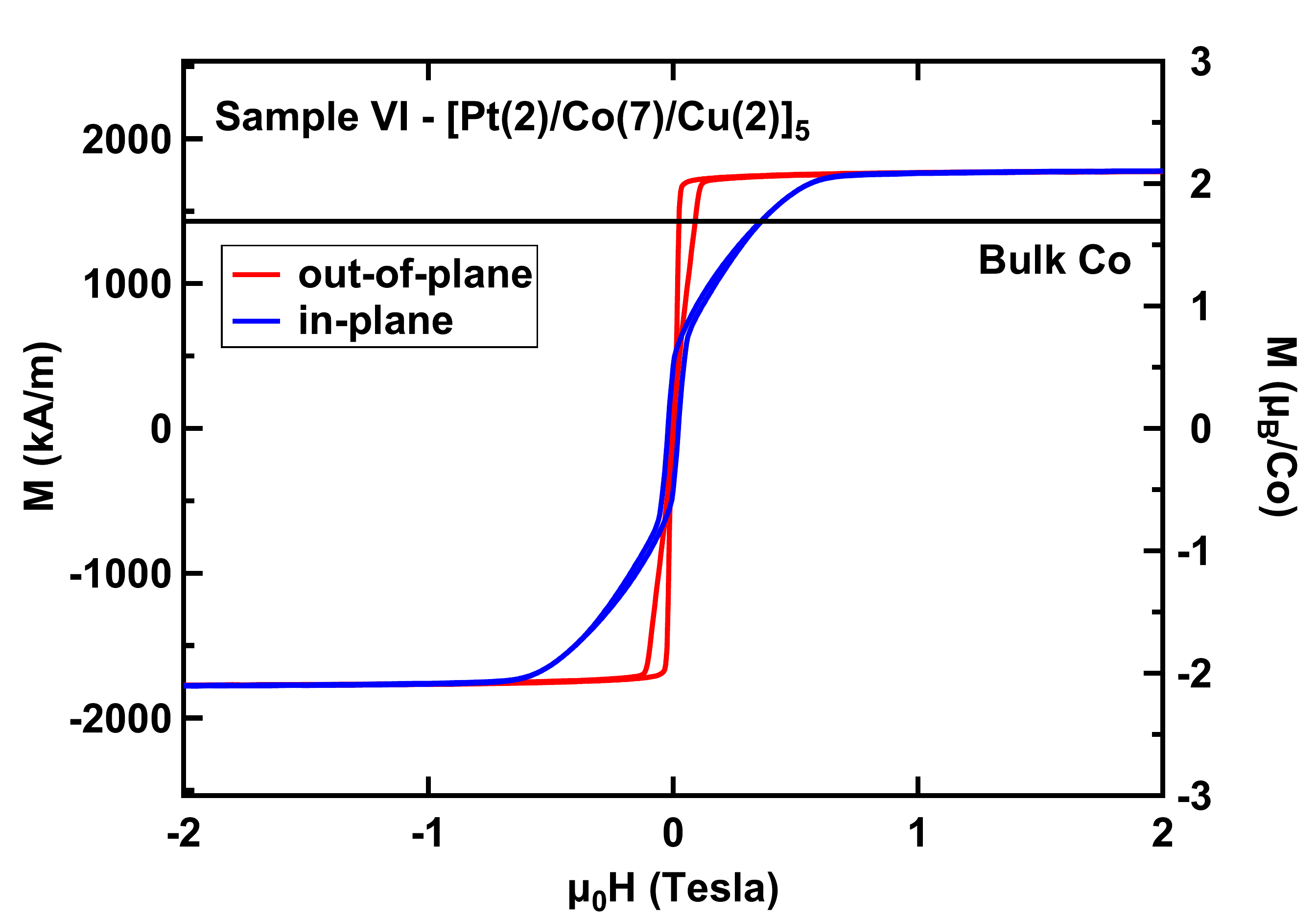}
    }
    \subfloat[\label{fig:SQUID_822}]{
       \includegraphics[width=0.48\textwidth]{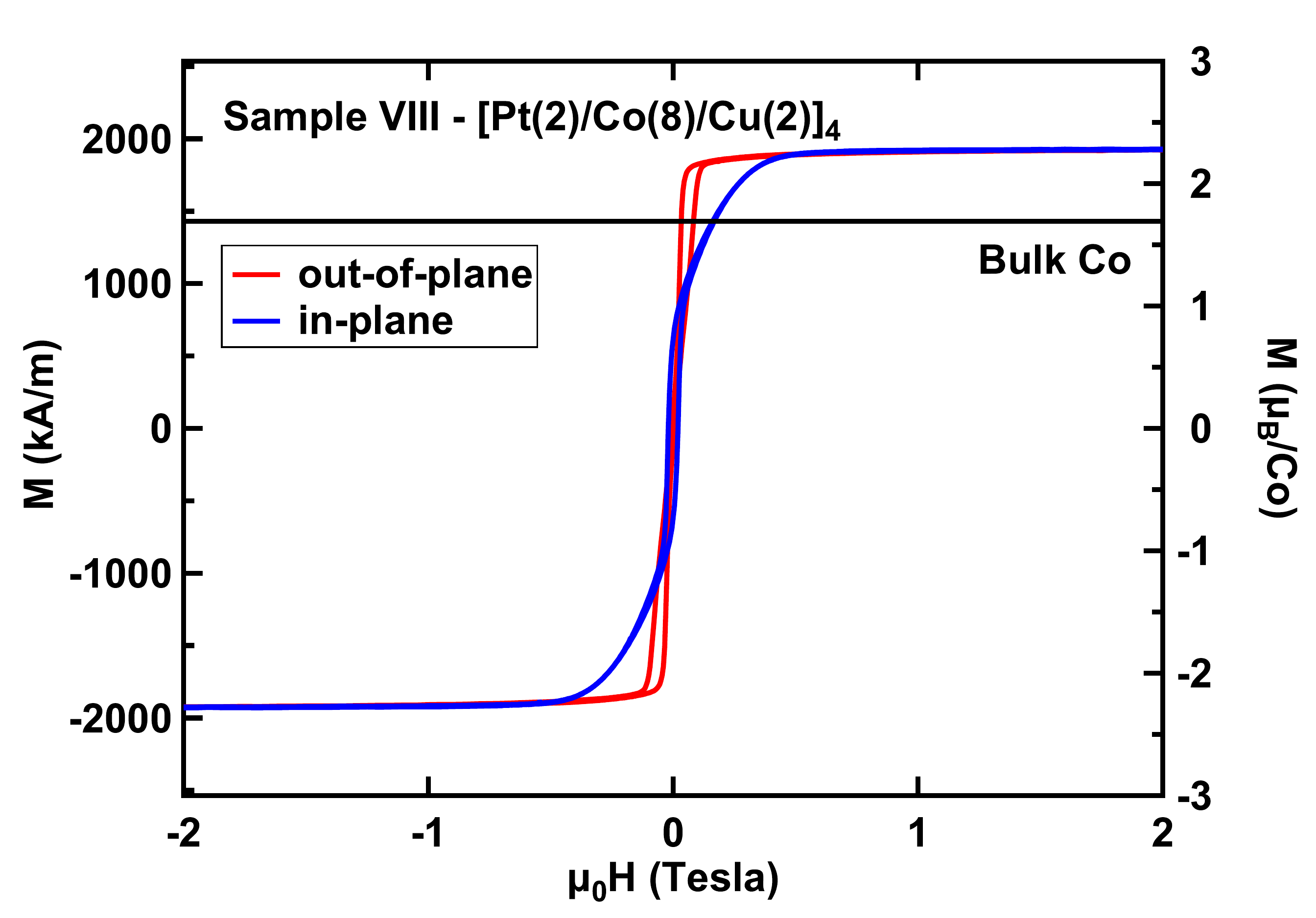}
    }\hfill
    \caption{\label{fig:Magnetic} SQUID data of 
    (a).  [Pt(2)/Co(7)/Cu(2)]$_5$ (sample VI) and (b). [Pt(2)/Co(8)/Cu(2)]$_4$ (sample VIII), respectively. 
    The black solid line represents 1430\,kA/m, which is the saturation magnetization of bulk Co. 
    } 
\end{figure*}

The out-of-plane (red) and in-plane (blue) hysteresis loops of [Pt(2)/Co(7)/Cu(2)]$_5$ (sample VI) and [Pt(2)/Co(8)/Cu(2)]$_4$ (sample VIII) are shown in Fig.~\ref{fig:SQUID_722} and ~\ref{fig:SQUID_822}, respectively.

\clearpage

\section{\NoCaseChange{LTEM details}}

\begin{figure*}[ht]
    \subfloat[\label{fig:LTEM_Sample_Tilt}]{
        \includegraphics[width=0.24\textwidth]{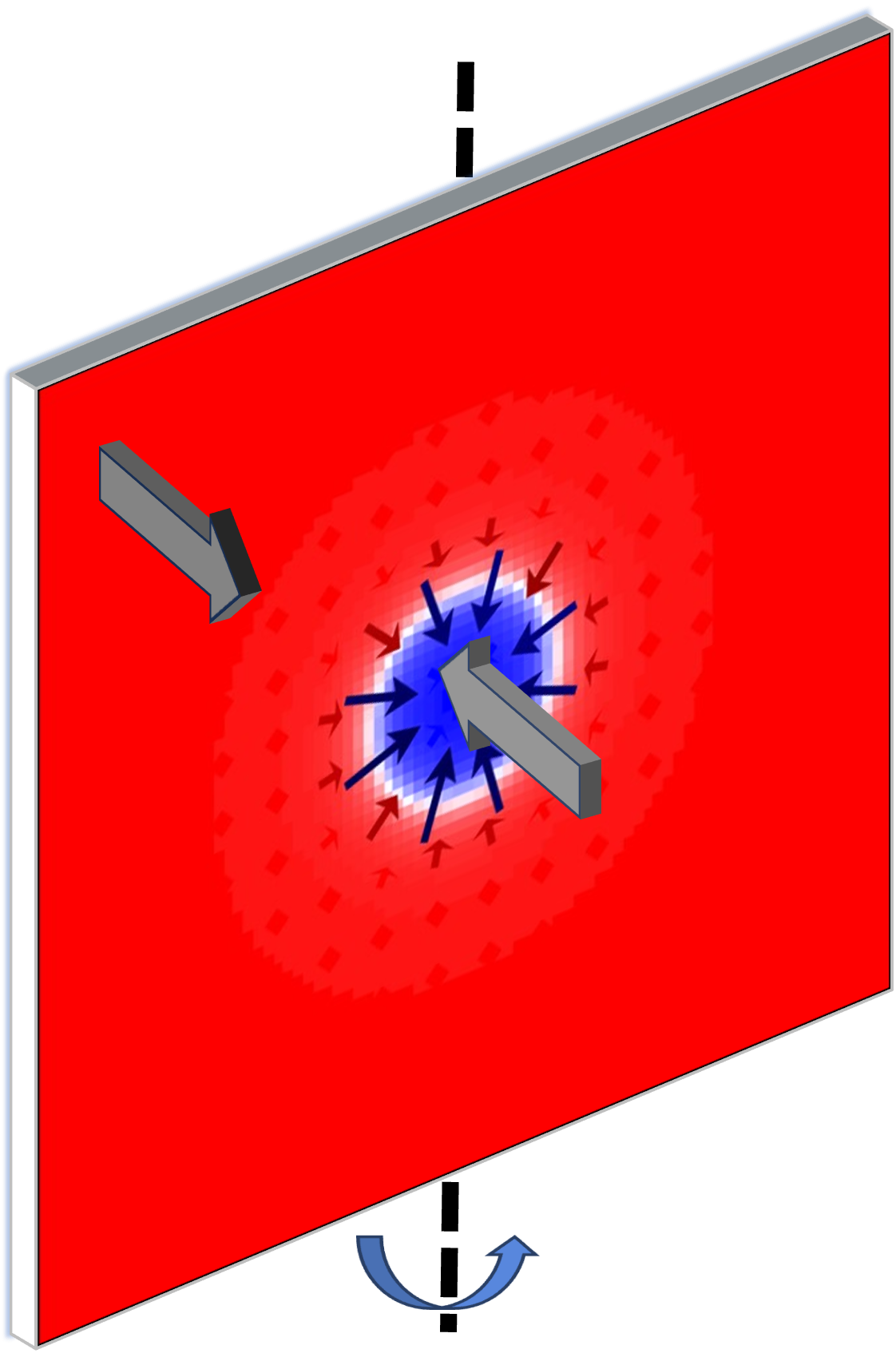}
        }\hspace{0.05\textwidth}
    \subfloat[\label{fig:LTEM_Simulate}]{
        \includegraphics[width=0.3\textwidth]{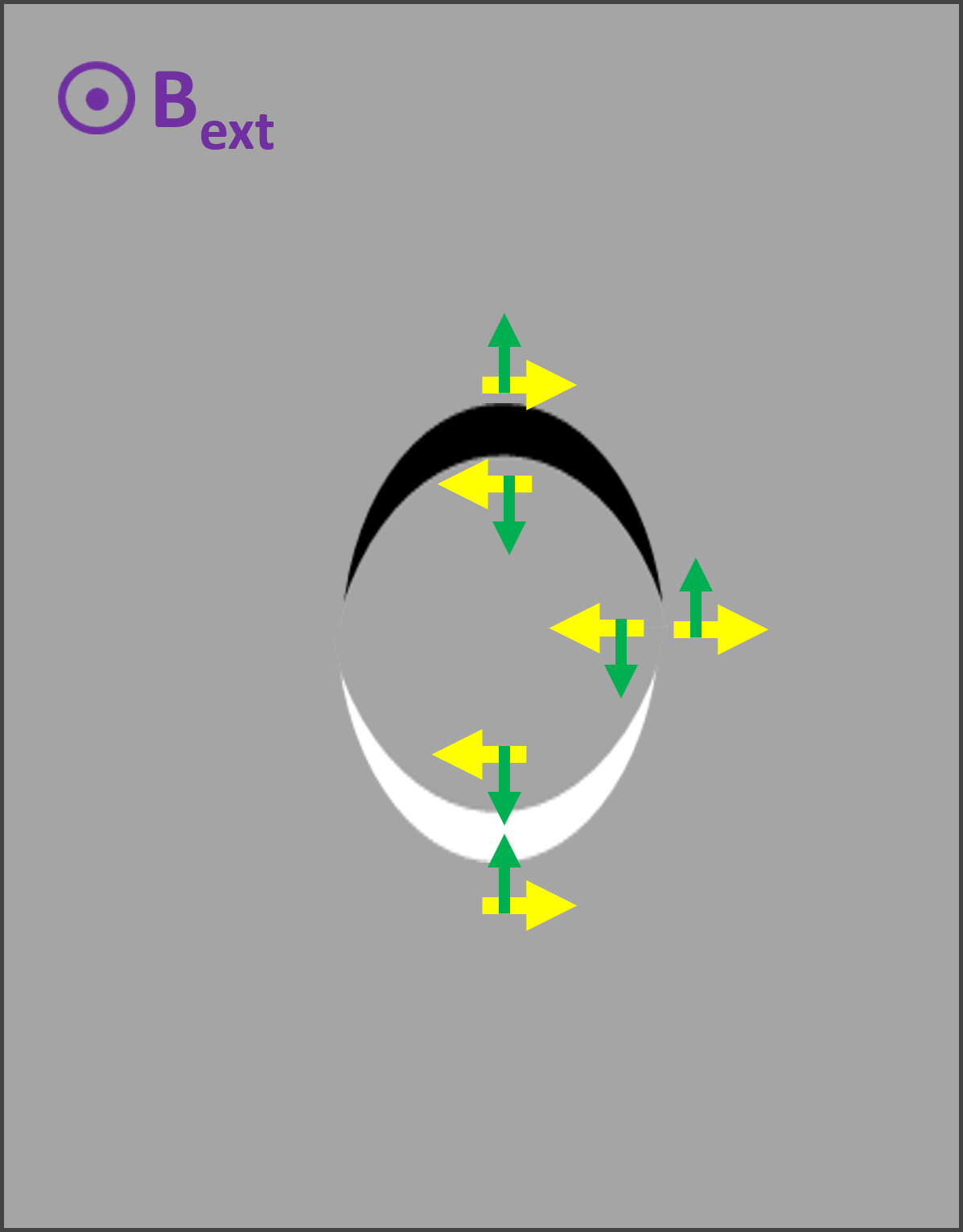}
        }
    \caption{\label{fig:LTEM_Tilt} (a). Schematic drawing of the LTEM experiment geometry, showing the tilt of the sample. The gray arrows represent the out-of-plane magnetization in the ``core'' region of the skyrmion and the outer zone. 
    (b) A simple drawing to illustrate the LTEM contrast of a N\'{e}el skyrmion. The external field $B_{ext}$ (purple) is out-of-plane. The yellow arrows represent the in-plane component of the magnetization, and the green arrows represent the direction of the Lorentz force.
    } 
\end{figure*}

Fig.~\ref{fig:LTEM_Sample_Tilt} shows the geometry of our LTEM measurement.
A skyrmion can be thought as being composed of an inner ``core'' region with magnetization antiparallel to the applied field, and an intermediate ``wall'' region where the magnetization rotates until it becomes parallel to the applied field in the ``outside'' region. 
In LTEM measurements, the electron beam is only deflected by the in-plane component of the magnetization.
Therefore, with normal incidence geometry, the inner core and the outer uniformly magnetized region do not deflect the electron beam.
Since the magnetization points along the radial direction in the ``wall" region of the N\'{e}el skyrmion, the Lorentz force points along the tangential direction.
Therefore, LTEM cannot image N\'{e}el skyrmions in the normal incidence geometry.
One way to observe N\'{e}el skyrmions by LTEM is to tilt the sample \cite{mcvitie2018transmission}, so that the magnetization has an in-plane component (the yellow arrows in Fig.~\ref{fig:LTEM_Simulate}).
In this case, the Lorentz force (the green arrows in Fig.~\ref{fig:LTEM_Simulate}) focuses (or defocuses) the electron beam at the top (bottom) of the skyrmion, leading to a positive and a negative lobe.
Because the contrast is generated by the magnetization of the core and outside regions and not the wall region, the contrast is determined by the sample tilt and not the chirality of the skyrmions.

\begin{figure*}[ht]
    \subfloat[\label{fig:LTEM_Linecuts}]{
        \includegraphics[width=0.45\textwidth]{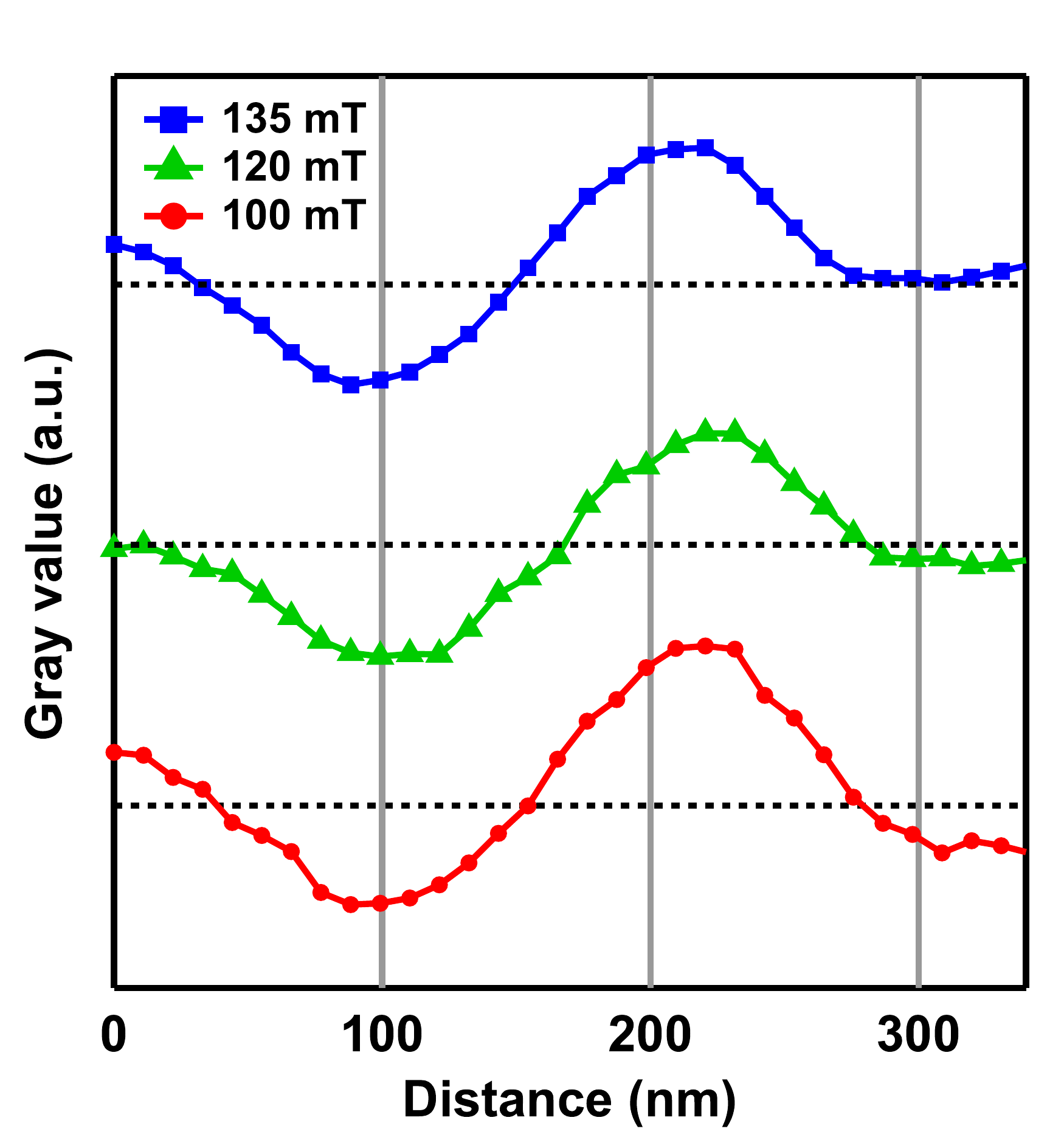}
        }
    \subfloat[\label{fig:LTEM_Hist}]{
        \includegraphics[width=0.45\textwidth]{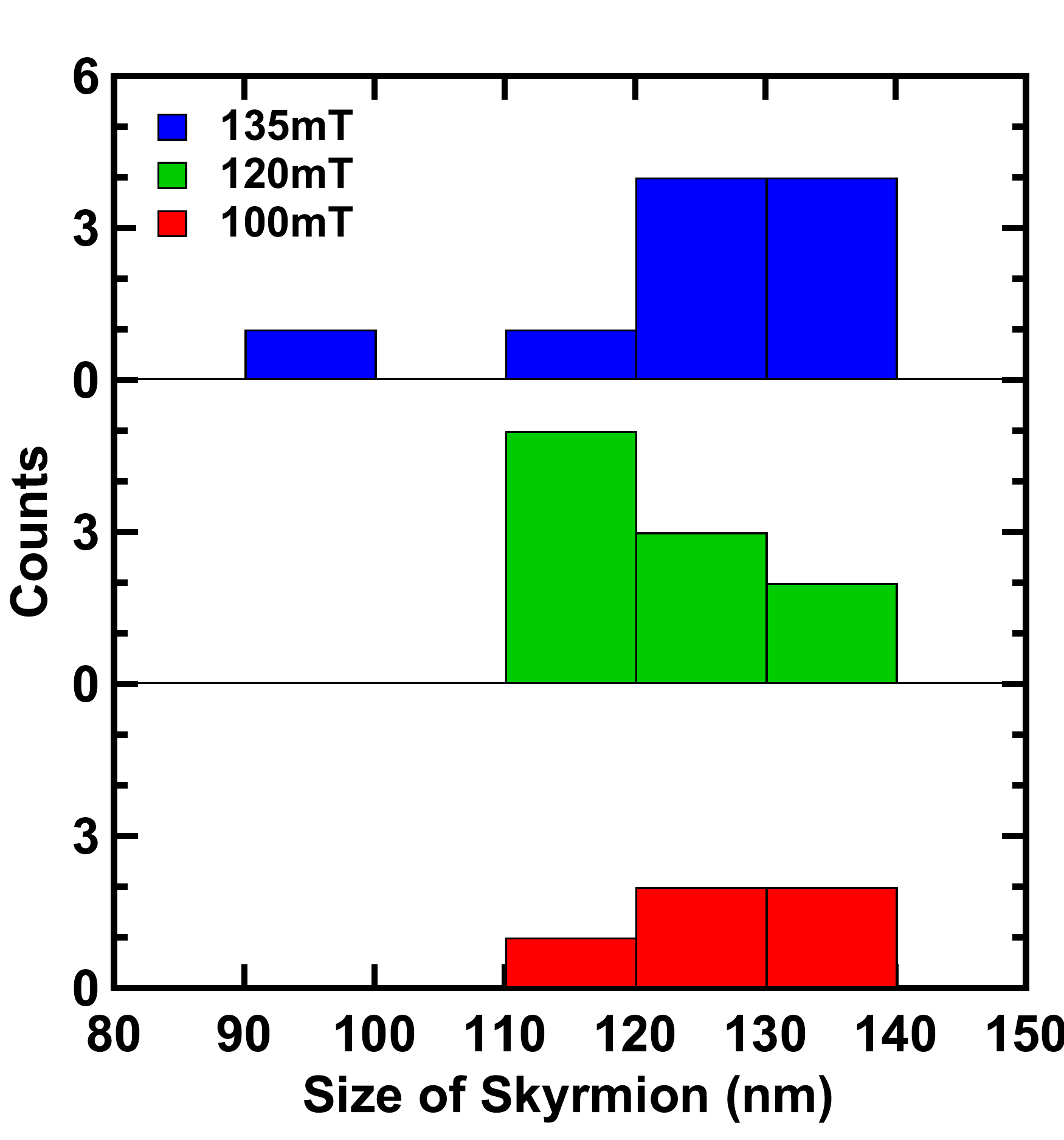}
        }
    \caption{\label{fig:LTEM_SkyrmionSize} Determination of average skyrmion sizes. 
    (a). Averaged line-cut profiles of multiple skyrmions at 100\,mT (red circles), 120\,mT (green squares), and 135\,mT (blue triangles) respectively, with 20$^{\circ}$ sample tilt.
    (b). Histograms of skyrmion sizes at 100\,mT (red), 120\,mT (green), and 135\,mT (blue), respectively.
    } 
\end{figure*}

Fig.~\ref{fig:LTEM_Linecuts} shows averaged line-cut profiles of skyrmions at 100\,mT (red circles), 120\,mT (green squares), and 135\,mT (blue triangles) respectively, with 20$^{\circ}$ sample tilt.
Each line-cut profile shown is an average of the line-cut profiles of multiple skyrmions at the same field.
The histograms in Fig.~\ref{fig:LTEM_Hist} represent the sizes of multiple skyrmions at each field, where the size of each skyrmion was determined from the maximum and minimum of its line-cut profile.
Analysis of the size distribution yields an average skyrmion size of $123\pm10$\,nm for 100\,mT, $118\pm10$\,nm for 120\,mT, and $122\pm12$\,nm for 135\,mT.
The error bars take into account the variation of the skyrmion size from the mean value (statistical error) and the uncertainty in determining the size of an individual skyrmion which is estimated to be $\pm\,6$\,nm (systematic error).
The average skyrmion sizes with error bars are plotted as a function of the applied field in Figure 6e of the main text.
We conclude that while the histograms might suggest a weak dependence of skyrmion size with the applied field, our analysis is unable to identify a field dependence of the skyrmion size within experimental uncertainties.

\clearpage

\section{\NoCaseChange{Additional MFM images}}

\begin{figure*}[ht]
        \includegraphics[width=\textwidth]{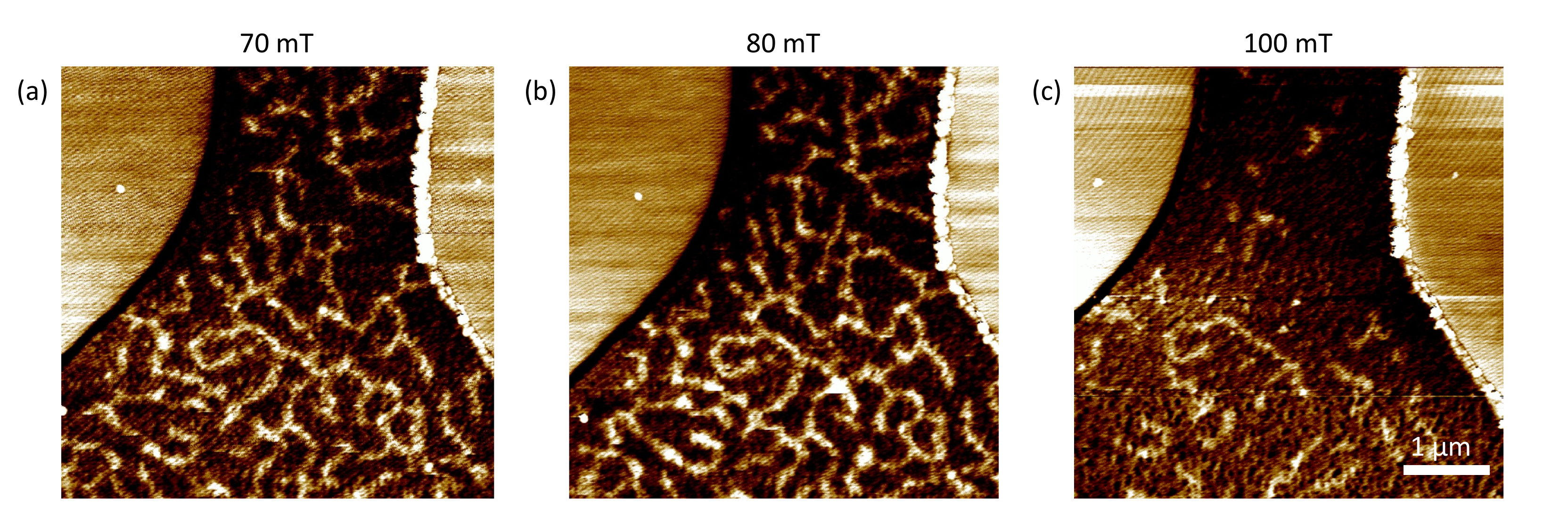}
    \caption{\label{fig:MFM_722} MFM images of [Pt(2)/Co(7)/Cu(2)]$_5$ sample with magnetic field ramped to (a) 70\,mT, (b) 80\,mT, and (c) 100\,mT.
    } 
\end{figure*}

We performed additional MFM measurements on [Pt(2)/Co(7)/Cu(2)]$_5$ sample, as shown in Fig.~\ref{fig:MFM_722}.
We started from zero field and ramped up the field up to 100\,mT gradually. 
Similar to [Pt(2)/Co(8)/Cu(2)]$_4$ sample sample, the magnetic textures were dominated by labyrinth domains at 70\,mT (Fig.~\ref{fig:MFM_722}a). 
At 80\,mT, the magnetic textures remained mostly unchanged as compared to 70\,mT (Fig.~\ref{fig:MFM_722}b).
However, as the field was ramped up to 100\,mT, a huge portion of labyrinth domains broke into skyrmions or disappeared, as shown in Fig.~\ref{fig:MFM_722}c.
This measurement demonstrates the existence of skyrmions in [Pt(2)/Co(7)/Cu(2)]$_5$ sample.

\begin{figure*}[ht]
        \includegraphics[width=0.5\textwidth]{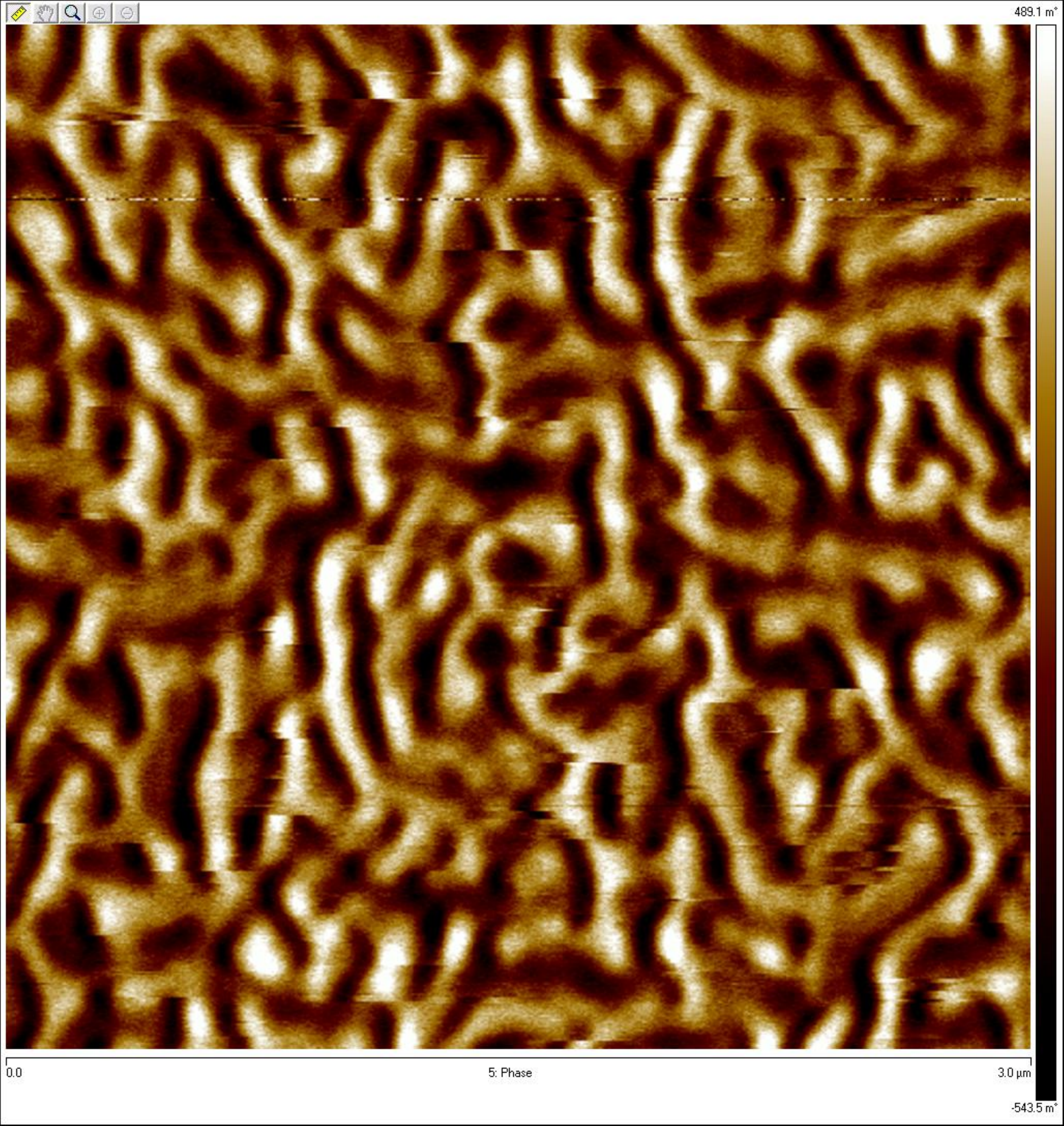}
    \caption{\label{fig:MFM_272_N7} MFM image of [Pt(2)/Co(7)/Cu(2)]$_7$ sample (sample VII) at zero magnetic field. The scan area is 3 microns by 3 microns.
    } 
\end{figure*}

To confirm that the [Pt(2)/Co(7)/Cu(2)]$_7$ sample (sample VII) is in the multidomain state, we measured MFM at zero field (Figure \ref{fig:MFM_272_N7}).

\clearpage

\section{\NoCaseChange Micromagnetic simulations}

We performed micromagnetic simulations of the [Pt(2)/Co(6)/Cu(2)]$_5$ sample on a slab consisting of 256$\times$256$\times$48 cells using MuMax3~\cite{vansteenkiste2014design}.
The size of each cell was 2$\times$2$\times$0.2\,nm$^3$, which allowed us to model [Pt(2)/Co(6)/Cu(2)]$_N$ as a $3N$-layer stack, with different saturation magnetization assigned to Co and Pt layers.
For the exchange stiffness $A_{ex}$ of cobalt, we used a typical value of 25 pJ/m~\cite{grimsditch_exchange_1997,ding_magnetic_2005, legrand_spatial_2022}. 
The interlayer exchange coupling was set to be $\sim$10\% of intralayer exchange coupling of Co~\cite{ounadjela1996dependence}.
The saturation magnetization and magnetic anisotropy were determined from SQUID magnetometry, shown in Figure 2b of the main text.
The saturation magnetization of Co layers was 1.4$\times$ 10$^3$\,kA/m, using the value of bulk cobalt~\cite{billas1994magnetism}. 
The magnetic moment of Pt was obtained from the following formula~\cite{bersweiler2016impact}:
\begin{equation}
\Delta M_{Co}\,t_{Co}^{total} = M_{Pt}\,t_{Pt}^{total} \label{eq1}
\end{equation}
By substituting $t_{Co}^{total}$ = 30 unit cells and $t_{Pt}^{total}$ = 10 unit cells into Eq.~\ref{eq1}, we got $M_{Pt}$ = 1.7$\times$ 10$^3$\,kA/m for micromagnetic simulation. 
Since the micromagnetic model required each atomic layer to have the same thickness (in this case 0.2\,nm), while in reality the Pt layers are thicker than Co layers, the $M_{Pt}$ was overestimated by $\sim$13\% here but is still within the uncertainty of experimentally measured $M_{Pt}$.
The in-plane and out-of-plane saturation fields in the SQUID measurement were used to find the effective anisotropy, which combines the magnetic anisotropy described by $K_u$ and the shape anisotropy of the sample. 
The uniaxial anisotropy was determined to be $K_u = 1.2\times 10^6$ J/m$^3$, following the discussion in section S2.

\begin{figure*}[ht]
        \includegraphics[width=0.2\textwidth]{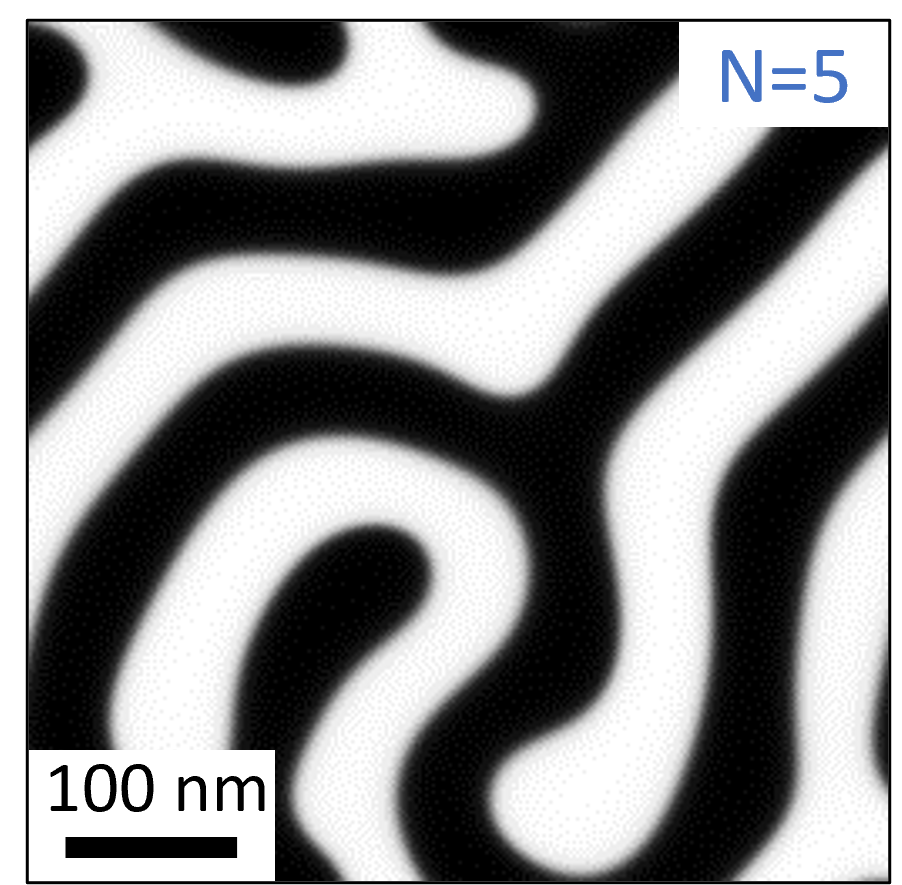}
    \caption{\label{fig:Simulation_622_5_0mT} Micromagnetic simulation results of [Pt(2)/Co(6)/Cu(2)]$_5$ relaxed from OOP polarized state at zero field, using exchange stiffness A$_{ex}$ = 25\,pJ/m.
    } 
\end{figure*}

\begin{figure*}[ht]
    \subfloat[\label{fig:Simulation_622_1_0mT_Aex=1}]{
        \includegraphics[width=0.18\textwidth]{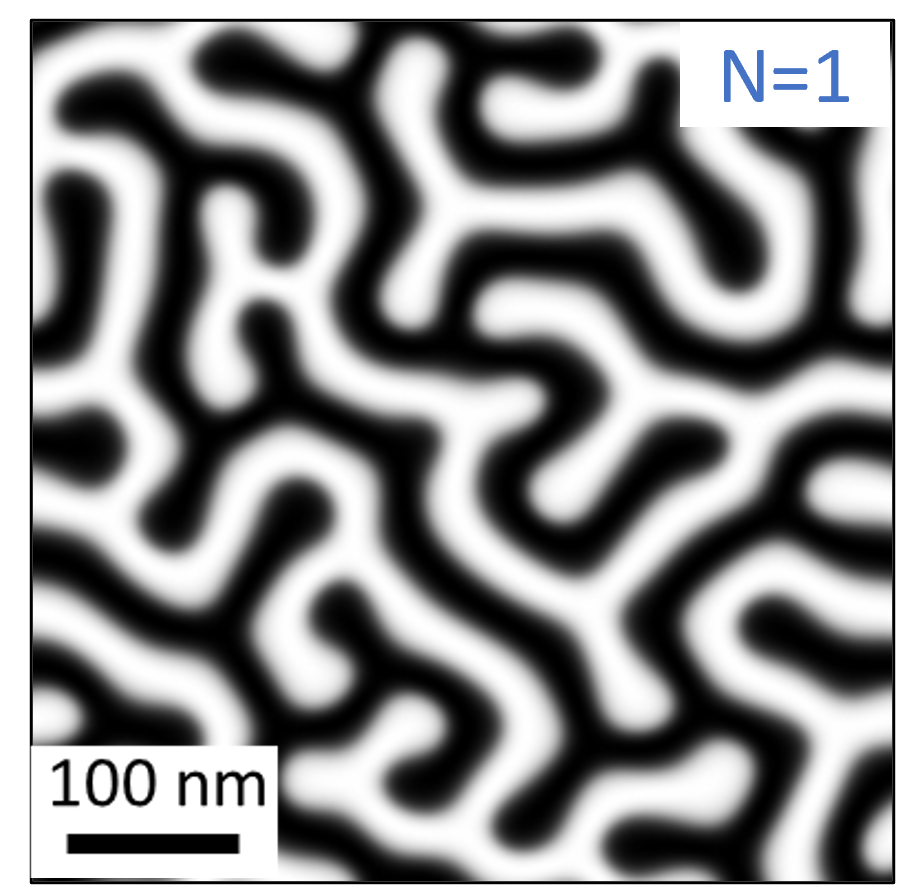}
    }
    \subfloat[\label{fig:Simulation_622_2_0mT_Aex=1}]{
        \includegraphics[width=0.18\textwidth]{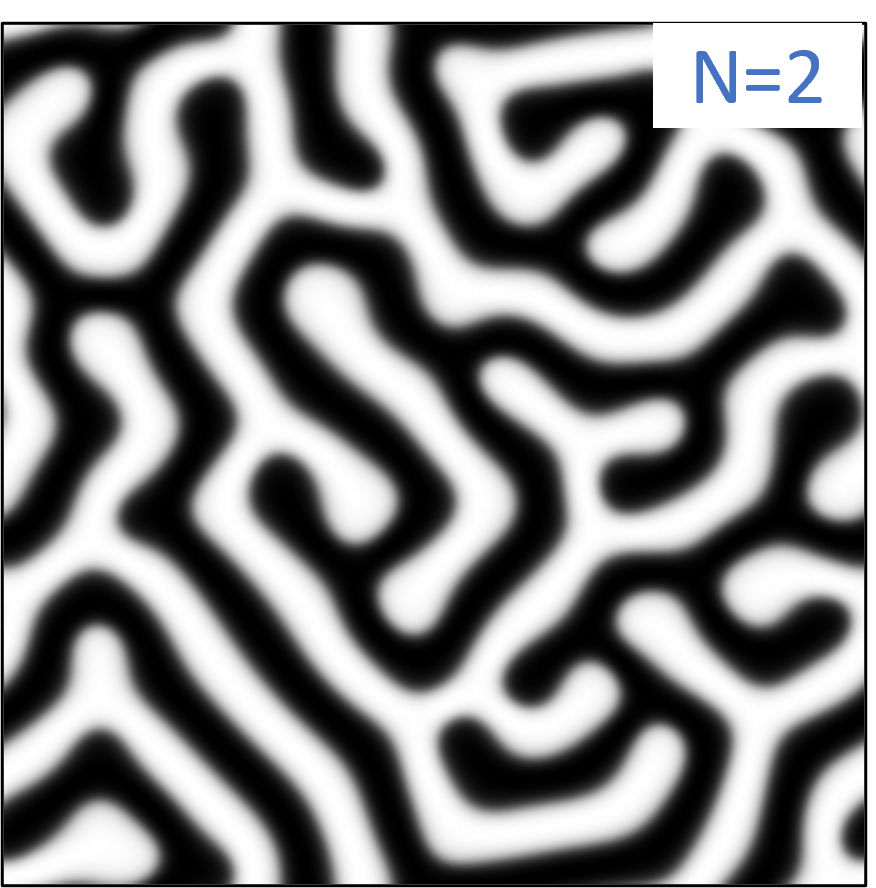}
    }
    \subfloat[\label{fig:Simulation_622_3_0mT_Aex=1}]{
        \includegraphics[width=0.18\textwidth]{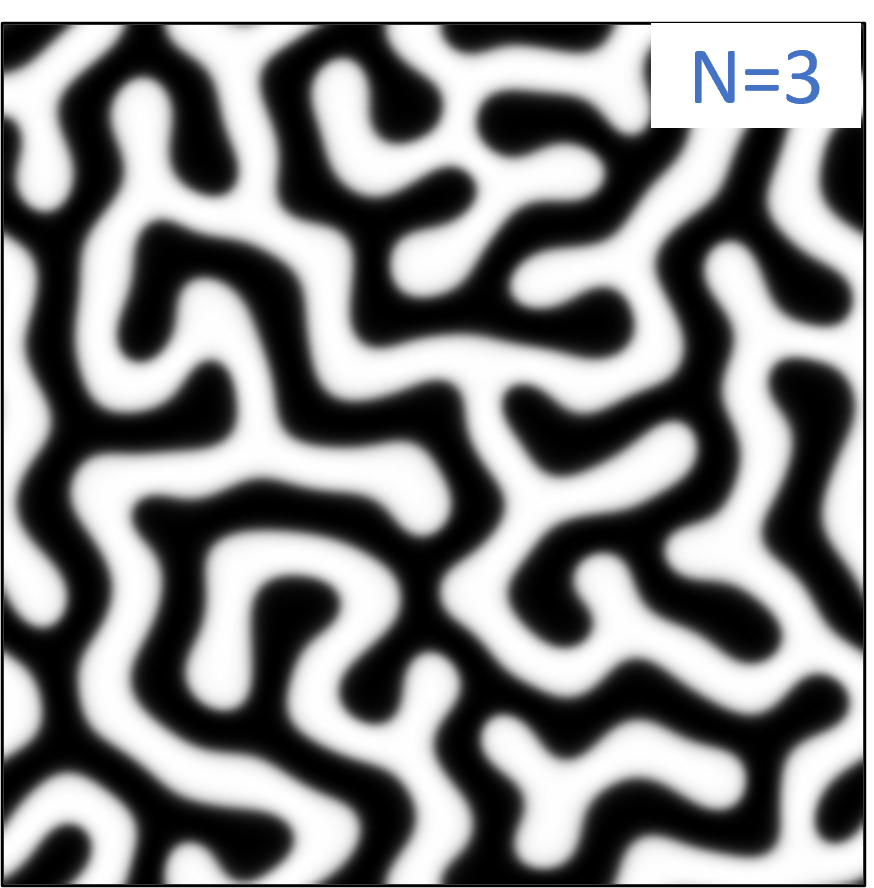}
    }
    \subfloat[\label{fig:Simulation_622_4_0mT_Aex=1}]{
        \includegraphics[width=0.18\textwidth]{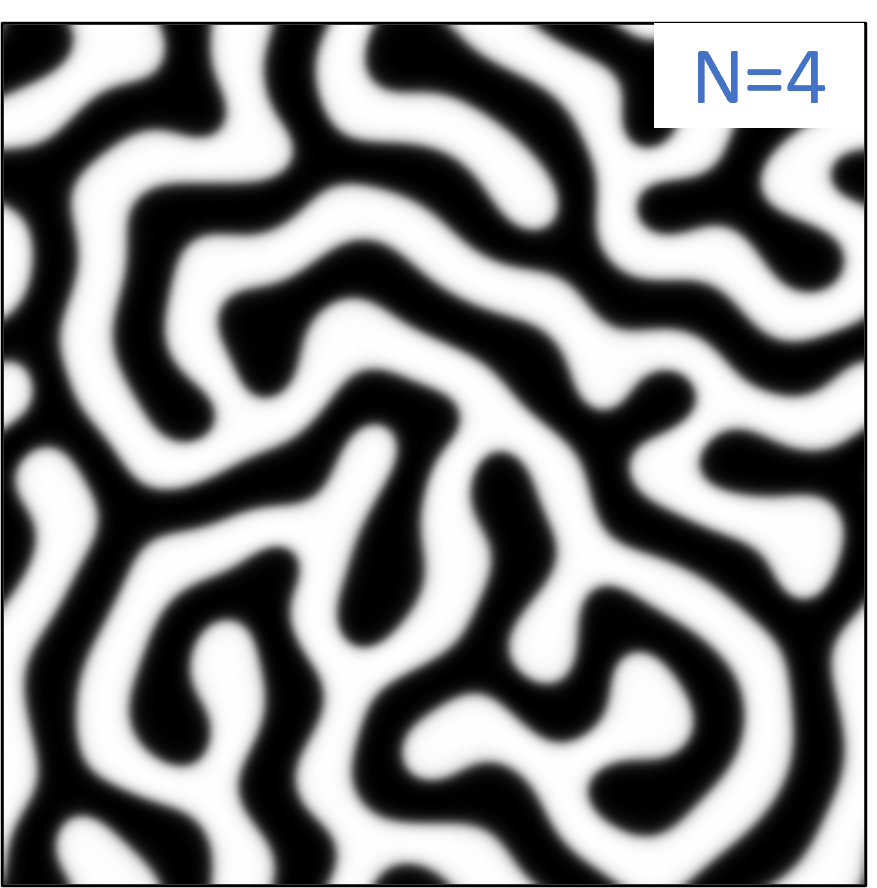}
    }
    \subfloat[\label{fig:Simulation_622_5_0mT_Aex=1}]{
        \includegraphics[width=0.18\textwidth]{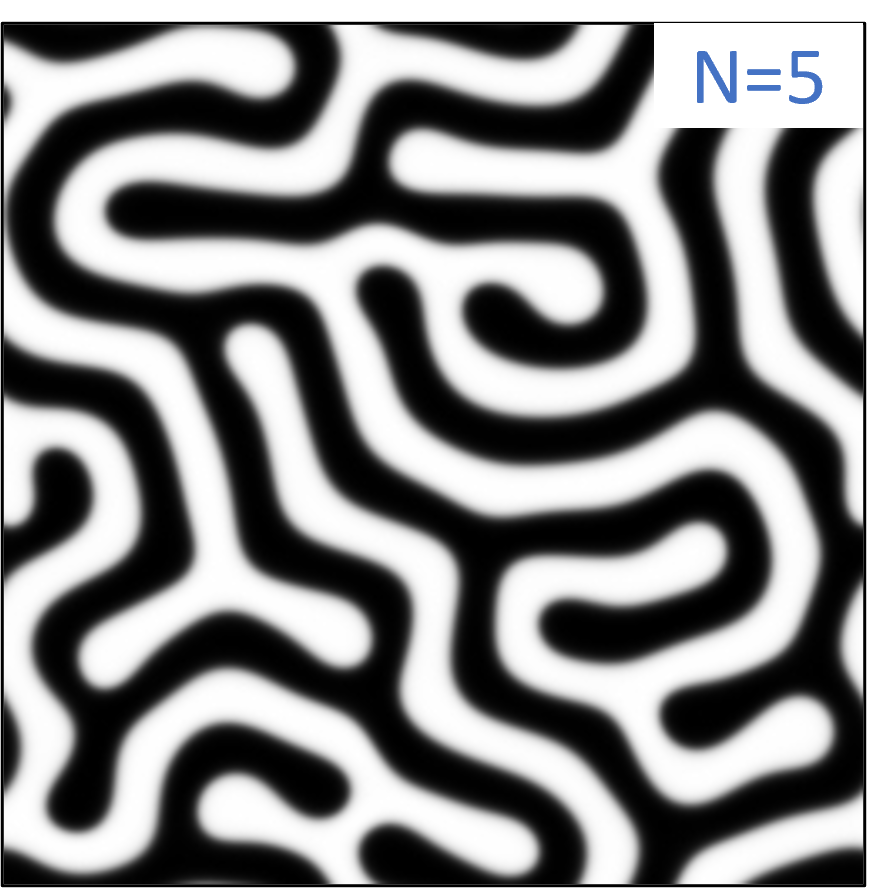}
    }
    \caption{\label{fig:Simulation_transition_Aex=1} Micromagnetic simulation results of [Pt(2)/Co(6)/Cu(2)]$_N$ relaxed from OOP polarized state at zero field, using exchange stiffness A$_{ex}$ = 10\,pJ/m. (a - e)  Simulation results of $N$ = 1 to $N$ = 5, respectively.
    } 
\end{figure*}

We first investigated the transition from uniformly OOP-magnetized states to multidomain states as the number of periods $N$ increases.
In these simulations, the initial state was along the OOP direction, with 1\% of the spin randomized, to avoid the possible unstable equilibrium states.
Then the system was allowed to relax at zero field.
For $N$ = 1 to 4, the magnetization remains uniformly magnetized at zero field (not shown here).
Meanwhile, for $N$ = 5, the uniformly magnetized state breaks into labyrinth domains at zero field, as shown in Fig.~\ref{fig:Simulation_622_5_0mT}.
However, with a smaller exchange stiffness $A_{ex}$ = 10\,pJ/m (which is another common value for Co~\cite{skomski_appendix_2008}), all of [Pt(2)/Co(6)/Cu(2)]$_N$ multilayers relax into multidomain states, as shown in Fig.~\ref{fig:Simulation_622_1_0mT_Aex=1} - \ref{fig:Simulation_622_4_0mT_Aex=1}.
This is because with smaller exchange stiffness, the exchange interaction is not enough to maintain a uniformly magnetized state, and therefore the dipolar interaction becomes dominant in the formation of multidomain states.
For N =5 case, fast Fourier transform (FFT) analysis yields an average labyrinth domain width of 120\,nm for $A_{ex}$ = 25\,pJ/m, and 55\,nm for $A_{ex}$ = 10\,pJ/m.
The domain width scales linearly with $A_{ex}$.

We next performed micromagnetic simulations to try to understand the skyrmion size as a function of the magnetic field and the strength of the DMI in the material.
The simulated skyrmion size was similar to the experimentally observed value.
We found for a given value of DMI, the skyrmion size decreases as the magnetic field increases, whereas for a given magnetic field, increasing the strength of the DMI increases the size of the skyrmion. The results are shown in Figure 6 of the main text.
To understand this, we use the skyrmion ansatz of Wang \emph{et al.}~\cite{wang2018theory} from their theory paper on skyrmion size. 
Our simulation results agree with their findings, that the skyrmion radius increases with DMI strength and decreases with magnetic field strength. 

\newpage

\bibliography{PtCoCu.bib}